\documentclass[11pt]{article}

\usepackage[a4paper,margin=1in]{geometry}
\usepackage{hyperref}
\usepackage{graphicx}
\usepackage{enumitem}
\usepackage{bm}
\usepackage{verbatim}
\usepackage{mathtools}
\usepackage{amssymb} 
\usepackage{physics}
\usepackage{mathrsfs}
\usepackage{exscale}
\usepackage{booktabs}
\usepackage{xcolor}

\graphicspath{{Figures/}}

\newcommand{\beq}{\begin{equation}}
\newcommand{\eeq}{\end{equation}}

\newcommand{\ee}{\mathrm{e}}

\newcommand{\OO}{\mathcal{O}}
\newcommand{\D}{\mathcal{D}}
\newcommand{\Order}[1]{\mathcal{O}\!\left(#1\right)}

\definecolor{darkred}{rgb}{0,0,0}
\definecolor{darkblue}{rgb}{0,0,0.7}

\begin{document}

\begin{center}
{\Large \textbf{Multi-Criticality and RG Topology in the Charge-Kondo-Breakdown Scenario in the Cuprates}}
\end{center}

\begin{center}
Stefan Kirchner$^{1\star}$ and Petr Jizba$^{2\ast}$
\end{center}

\begin{center}
{\bf 1} Department of Electrophysics, National Yang Ming Chiao Tung University, Hsinchu 30010, Taiwan\\
{\bf 2} FNSPE, Czech Technical University in Prague, B\v{r}ehov\'{a} 7, 115 19 Praha 1, Czech Republic\\
$^\star$ \href{mailto:kirchner@nycu.edu.tw}{kirchner@nycu.edu.tw}~;~~~
$^\ast$ \href{mailto:petr.jizba@fjfi.cvut.cz}{petr.jizba@fjfi.cvut.cz}
\end{center}

\begin{center}
\today
\end{center}

\begin{abstract}
\noindent\bfseries
In this paper, we examine the dynamical charge-Kondo-breakdown scenario proposed for cuprate superconductors within the  perspective  of re\-nor\-mal\-i\-za\-tion-group (RG) topology. By analyzing the coupled RG flow equations governing the effective low-energy theory, we determine the fixed-point structure, stability properties, and global organization of the flow.  We find that the putative finite-coupling interacting fixed point is unstable against perturbations transverse to an invariant critical manifold. As a result, generic RG trajectories exhibit runaway behavior.
To elucidate the global structure of the theory, we combine analytical solutions of the flow equations with numerical phase portraits and Poincaré compactification. This analysis reveals that the interacting fixed point exhibits a marginally relevant instability, generating an exponentially large but finite crossover scale. The resulting flow topology closely resembles anisotropy-driven runaway flows encountered in fluctuation-induced weakly first-order transitions. Within this framework, the apparent quantum-critical regime can be understood as an extended crossover controlled by a near-critical fixed point, rather than by asymptotic scale invariance. The results obtained indicate that the existence of extended scaling behavior does not, by itself, imply the presence of a stable interacting quantum critical state. More generally, our analysis demonstrates how RG topology can provide a powerful diagnostic for distinguishing genuine criticality from pseudo-critical behavior in theories of strongly correlated quantum matter.
\end{abstract}

\tableofcontents

\section{Introduction and Background}
\label{intro}

The discovery of high-temperature superconductivity in the cuprates revealed a broad region of the phase diagram whose normal-state properties depart strongly from the predictions of Landau Fermi-liquid theory, \textit{i.e.}, the standard model of the metallic state.  A prominent example is the strange-metal regime, in which the electrical resistivity exhibits an approximately linear temperature dependence over an exceptionally wide temperature range, extending in some materials down to temperatures near and above the superconducting transition temperature~\cite{Keimer.15,Varma.89,Legros.19}. Concomitantly, a growing body of experimental evidence indicates that transport relaxation rates approach the so-called Planckian limit, $\hbar/\tau \sim k_B T$,  suggesting that the underlying dynamics may be governed by strongly interacting degrees of freedom for which the quasiparticle description ceases to be applicable~\cite{Bruin.13,Zaanen.04,Hartnoll.15}.
	
Despite several decades of investigation, no consensus has emerged regarding the microscopic origin of strange metallicity. One influential line of thought attributes the observed scaling behavior to the proximity of a quantum critical point (QCP), where fluctuations associated with a zero-temperature phase transition generate non-Fermi-liquid dynamics over an extended finite-temperature regime~\cite{Sachdev.11,Lohneysen.07}. More recently, holographic approaches based on gauge--gravity duality have provided alternative phenomenological frameworks capable of producing scale-invariant transport and Planckian relaxation~\cite{Hartnoll.18,Zaanen.15}. At the same time, a variety of microscopic mechanisms have been proposed in which Fermi-liquid behavior is destabilized via Kondo breakdown, electron fractionalization, emergent gauge fields, or locally critical degrees of freedom~\cite{Si.01,Coleman.01,Senthil.04,Si.14}.
This has been particularly successful in rare earth-based intermetallic compounds, where experimental evidence supports the existence of critical Kondo destruction~\cite{Friedemann.10,Kirchner.20}.
	
Within this broader context, Chang~\emph{et al.} have recently proposed a dynamical charge-Kondo breakdown scenario for cuprate superconductors~\cite{YY-Chang.25}. Their construction is motivated by the idea that a localized-to-delocalized transition of charge degrees of freedom may generate an interacting quantum-critical regime, potentially accounting for both $\omega/T$ scaling and Planckian dissipation. Starting from the $t$-$J$ model --- a standard framework for describing the low-energy physics of cuprates --- they implement a coarse-graining procedure that leads to an effective renormalization-group (RG) flow. This flow is argued to contain a quantum critical fixed point, which controls four phases: a Fermi liquid, a superconducting state, a pseudogap regime, and a strange-metal phase. Given that this approach aims to provide a microscopic route to strange metallicity without invoking holographic duality techniques, e.g. along the lines of~\cite{Zaanen.15,LUCAS2015}, it is important to understand in detail the structure and assumptions underlying the resulting RG flow equations and their connection to the proposed phase diagram.
	
The structure of the RG flow is central to the analysis of quantum critical behavior and can be studied independently of the microscopic assumptions that lead to a given set of scaling equations. The existence of fixed points alone is not sufficient to determine the nature of the critical behavior. One must also understand their stability properties, the dimensionality of the associated critical manifolds, and the global structure of the flow in coupling space. These features determine whether scaling emerges asymptotically, whether it requires fine-tuning of control parameters, and whether an apparently critical regime is governed by a genuine continuous phase transition or instead represents a crossover with an extended but finite scaling regime.
	
There are many examples where this distinction is essential. The Coleman--Weinberg mechanism~\cite{Coleman.73} and the work of Halperin, Lubensky, and Ma~\cite{Halperin.74} show that fixed points which appear critical at a given level of approximation may become destabilized upon inclusion of additional directions in coupling space, resulting in runaway flows and fluctuation-induced first-order transitions. Related behavior is found in models with cubic anisotropy, where interacting fixed points exist but are unstable against symmetry-breaking perturbations~\cite{Aharony.73,Bruce.75,Rudnick.78}.
This motivates an examination of the RG flow directly in coupling space. In what follows, we distinguish between the physical tuning parameter that accesses criticality and the interaction couplings appearing in the RG equations, the latter being the primary object of our stability analysis.

Rather than addressing the broader phenomenological viability of the dynamical charge-Kondo breakdown scenario, which certainly exhibits a number of appealing features in its own right, we focus on the mathematical structure of the associated RG equations. We do not consider the derivation of these equations from the microscopic $t$-$J$ model, {which constitutes one of the most significant and technically interesting aspects of Ref.~\cite{YY-Chang.25}.} Instead, our emphasis is more modest and concerns the analysis and interpretation of the resulting RG flow in theory space. In particular, we determine the fixed-point structure, analyze the local stability properties, and obtain both analytical and numerical solutions of the flow equations. The global phase portrait is further examined using Poincar\'e compactification.
This analysis reveals a flow topology that differs in important respects from the scenario proposed in Ref.~\cite{YY-Chang.25} {and depends on the specific form of the RG flow equations under consideration. In the main text, we focus on the original set of equations in Ref.~\cite{YY-Chang.25}, prior to rescaling of the coupling constants, which we refer to as \textit{the RG equations}. The rescaled formulation is discussed separately in Appendix C.}
{For the RG equations, we conclude that} the interacting fixed point possesses a marginally relevant transverse direction, which generates runaway RG trajectories of the type familiar from fluctuation-induced weakly first-order transitions. Accordingly, any extended scaling regime should be interpreted as pseudo-critical rather than asymptotically scale invariant.
{The rescaled equations, in contrast, admit a multi-critical point that acts as a UV fixed point. We thus use the term \textit{multi-critical} in a broader sense, encompassing both conventional multi-critical behavior and pseudo-critical runaway flows.}

The remainder of this paper is organized as follows. In Section~\ref{RGflowDiagram}, we introduce the RG equations underlying the charge-Kondo breakdown scenario and discuss their fixed-point structure. Section~\ref{sec:omega-over-T} analyzes the local stability of these fixed points and constructs the associated phase portrait. In Section~\ref{sec:solution}, we solve the RG equations both analytically and numerically, and study the global structure of the flow using Poincar\'{e} compactification, which allows us to characterize the asymptotic behavior of trajectories in coupling space. Section~\ref{runaway} places the resulting flow structure in the broader context of anisotropy-driven runaway flows and fluctuation-induced weakly first-order transitions, and discusses the implications for the proposed quantum-critical regime. Finally, the conclusions are summarized in Section~\ref{Sec:conclusion}.
For the reader's convenience, several technical and conceptual issues related to stability and topological invariants associated with two-dimensional RG flows are
relegated to {four} appendices. {This also includes a discussion of the behavior of RG flow equations under coupling constant rescalings and field redefinitions.}
	
\section{Information encoded in an RG flow diagram}
\label{RGflowDiagram}
	
An RG flow diagram provides a graphical representation of the evolution of a theory under coarse-graining transformations. As the description scale is varied from short distances, also called the ultraviolet (UV) limit, to long distances, i.e., the infrared (IR) limit, the flow identifies the couplings that control the long-distance behavior of the system.
	
The coordinates of the flow diagram are the coupling constants $g_i$. The RG flow is generated by the beta functions, which define a vector field on the space of couplings. In the standard quantum field theory convention, one introduces a running momentum scale $\mu$ (with $\mu \le \Lambda$, where $\Lambda$ is the UV cutoff), and defines
\begin{eqnarray}
	\beta_i(\mathbf{g}) \ \equiv \ - \frac{d g_i}{d \ln(\mu/\Lambda)} \ =\  - \frac{d g_i}{d \ln \mu}\, .
\end{eqnarray}
This definition makes explicit that the beta functions encode the response of the couplings under changes of renormalization scale. The minus sign is chosen so that the RG flow points towards the infrared as \(\mu\) is decreased.  The latter is opposite to the convention used in particle physics.
	
The integral curves of the vector field $\beta_i(\mathbf{g})$ are the RG trajectories, describing how the couplings vary under successive coarse-graining transformations. Each point in coupling space corresponds to a particular theory, while an RG trajectory represents its continuous deformation under changes of scale. Since the RG transformation is conventionally implemented by integrating out short-distance (high-energy) degrees of freedom, it is natural to introduce an RG ``time''
\begin{eqnarray}
	\ell  \ \equiv \  \ln(\Lambda/\mu)\, ,
\end{eqnarray}
which increases as one flows towards the infrared. In terms of this variable, the flow equations take the canonical form
\begin{eqnarray}
	\frac{d g_i}{d \ell} \ = \  \beta_i(\mathbf{g})\, ,
\end{eqnarray}
so that the beta functions directly generate the RG trajectories as integral curves, and the flow is oriented towards increasing length scales.

Fixed points $\mathbf{g}^*$ are defined as the zeros of the beta functions, i.e., when $\beta_i(\mathbf{g}^*) = 0$,
so that the couplings cease to flow under RG transformations. This implies that the theory becomes scale-invariant at $\mathbf{g}^\ast$, since the RG vector field vanishes and there is no distinguished scale dependence present. The nature of a fixed point is determined by linearising the RG flow in its vicinity. Writing $g_i = g_i^\ast + \delta g_i$, one obtains (see also Appendix~A)
\begin{eqnarray}
	\frac{d}{d\ell} \ \! \delta g_i \ = \  M_{ij}\,\delta g_j\, ,
	\qquad
	M_{ij} \ \equiv \  \left.\frac{\partial \beta_i}{\partial g_j}\right|_{g=g^\ast}\, ,
	\label{II.4.rf}
\end{eqnarray}
so that the stability properties are governed by the eigenvalues of the {\em stability matrix} $M_{ij}$. A fixed point is called \emph{stable} (or attractive) if all perturbations which are not forbidden by symmetries are driven back towards $\mathbf{g}^\ast$ under RG flow, i.e. if the corresponding eigenvalues lead to decay of $\delta g_i$ as $\ell \to \infty$. Conversely, if there exist directions in coupling space for which perturbations grow under RG, the fixed point is \emph{unstable} (or repulsive).
	
Stable fixed points describe IR phases of matter: generic RG trajectories in their basin of attraction flow towards the same limiting point, and therefore exhibit identical long-distance behavior. Such sets of theories define a {\em universality class}. Unstable fixed points, by contrast, are typically associated with critical points separating distinct phases. Reaching them requires tuning the relevant couplings so as to eliminate all directions along which the flow is repelled; they therefore have co-dimension equal to the number of relevant operators. Here, by a {\em relevant operator} we mean an operator whose coupling increases under coarse-graining.

\subsection{Scaling Behavior}
%
As already mentioned, near a fixed point, the flow is governed by the linearized equations~(\ref{II.4.rf}). The eigenvalues $\lambda_a$ of the matrix $M_{ij}$ determine the nature of the corresponding directions in coupling space  (see also Appendix~A): If $\mathrm{Re}(\lambda_{1,2})  > 0$, perturbations grow as the IR limit is approached, causing trajectories to be repelled from the fixed point. Such a direction is called \textit{relevant}.   Conversely, if $\mathrm{Re}(\lambda_{1,2}) < 0$, perturbations decay, and trajectories are attracted towards the fixed point. The ensuing direction is called \textit{irrelevant}.  In the \textit{marginal} case, where $\mathrm{Re}(\lambda_{1}) = 0$ and/or $\mathrm{Re}(\lambda_{2}) = 0$, the corresponding coupling(s) remain approximately constant at linear order, and a higher-order analysis is required.
	
Thus, to interpret a flow diagram,  one follows the arrows towards larger scales (or lower energies) to identify the stable IR behavior.  If a trajectory is pushed away from a fixed point, the corresponding direction is relevant and determines the critical behavior of the system.
%
\begin{figure}[h!t]
	\centering \includegraphics[scale=0.2]{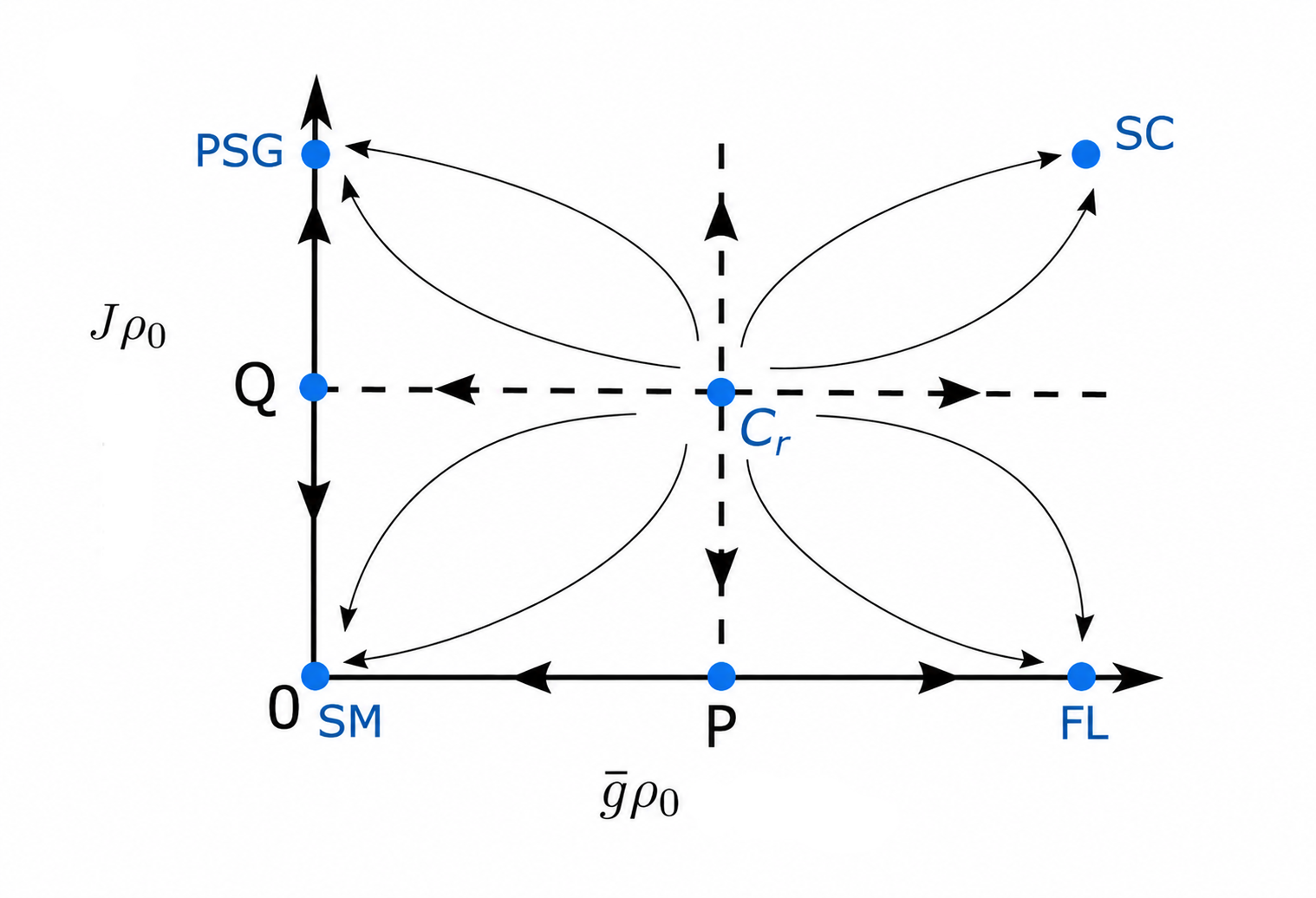}
	\caption{Schematic renormalization-group flow diagram adapted from Fig. 3(a) of Ref.~\cite{YY-Chang.25}. The central fixed point $C_r$
	is depicted as unstable along both coupling directions, corresponding to a total-repeller fixed point. Such a topology implies that all perturbations grow under coarse graining and has important consequences for the robustness of quantum-critical scaling and $\omega/T$ collapse.}
	\label{fig:RGflow}
\end{figure}
%
With this in mind, we now turn to the RG flow diagram obtained by Chang et al.~\cite{YY-Chang.25} for the $t$-$J$ model, a commonly used proxy for the low-energy physics of the Hubbard model near half-filling. The RG flow diagram of Ref.~\cite{YY-Chang.25} is reproduced in Fig.~\ref{fig:RGflow}.
The diagram features several fixed points at non-negative coupling, indicated by the blue dots.
Its central feature is the intermediate coupling fixed point labeled $C_r$. This fixed point is relevant in both directions, i.e. fully unstable.  We will refer to this feature as a ``total-repeller'' critical point. {Critical points with more than one relevant direction are commonly associated with multi-critical behavior.}
	
\subsection{Physical Nature and Ramifications of a Total-Repeller QCP}
	
A QCP whose independent non-redundant scaling fields are all relevant corresponds to a fixed point with no infrared-attractive scaling directions in the physical subspace under consideration. Under coarse graining, any infinitesimal displacement away from the fixed point grows along the corresponding scaling direction, so that nearby trajectories are expelled from its neighborhood as the flow proceeds toward the infrared. Such a point is therefore not a conventional codimension-one critical point, but rather a high-codimension unstable fixed point whose critical behavior can be accessed only when all relevant scaling fields are tuned sufficiently close to zero.
	
In the convention used here, the RG flow is defined by coarse graining, or equivalently by increasing the length scale $b=e^\ell$. Near a source-like QCP, all relevant perturbations grow under flow toward the infrared. Therefore, trajectories generically move away from the fixed point as $b$ increases. Conversely, the fixed point is locally attractive only under the inverse flow, namely when the RG equations are evolved backward toward the ultraviolet. Such a fixed point may therefore organize the neighboring basins of attraction of several infrared phases, but it is not an infrared-stable critical point. In contrast, an ordinary one-parameter QCP is usually a saddle in theory space: it has one relevant direction, corresponding to the experimentally tuned control parameter, and irrelevant directions along which the critical surface is attractive.
	
This distinction has direct consequences for scaling. Let $u_i$ denote the independent scaling fields near the fixed point, with associated RG eigenvalues $y_i$. In the strict total-repeller case, all $y_i>0$. Assuming hyperscaling, the singular part of the free-energy density satisfies the scaling form
\begin{eqnarray}
	f_s(u_1,\dots,u_N,T)
	\ =\
	b^{-(d+z)} \Phi \left( u_1 b^{y_1}, \dots,
	u_N b^{y_N}, T b^z \right),
\end{eqnarray}
where $d$ denotes the spatial dimension and $z$  the dynamical critical exponent. The exponent $z$ is conventionally non-negative and is positive for ordinary dynamical scaling.
In situations where hyperscaling is violated, the prefactor should be replaced by $b^{-(d+z-\theta)}$, where $\theta$ is the hyperscaling-violation exponent.
	
Choosing $b$ such that $T b^z=1$ gives the finite-temperature scaling form
\begin{eqnarray}
	f_s
	\ =\
	T^{(d+z)/z} \ \!\tilde{\Phi} \!\left( \frac{u_1}{T^{y_1/z}}, \dots,
	\frac{u_N}{T^{y_N/z}} \right),
\end{eqnarray}
again assuming hyperscaling. Thus, in a genuine total-repeller problem, finite-temperature observables are functions of several independent scaling variables. For example,
\begin{eqnarray}
	c_V
	\ =\
	-T \left. \frac{\partial^2 f_s}{\partial T^2} \right|_{u_i,V}
\end{eqnarray}
and susceptibilities obtained by differentiating $f_s$ with respect to appropriate source fields will generally depend on all ratios $u_i/T^{y_i/z}$. As a result, single-parameter scaling is not expected unless the physical trajectory in parameter space is restricted to a special direction, or unless all but one relevant scaling field are negligibly small over the experimentally accessible window.
	
Equivalently, each relevant scaling field generates an associated crossover scale
\begin{eqnarray}
	\Delta_i \ \sim \  |u_i|^{z/y_i}\, ,
\end{eqnarray}
or, equivalently, along the corresponding scaling direction, a crossover length scale
\begin{eqnarray}
	\xi_i \ \sim \  |u_i|^{-1/y_i}\, .
\end{eqnarray}
The quantum-critical regime is reached only when the temperature constitutes the largest infrared energy scale induced by the relevant perturbations, namely
\begin{eqnarray}
	k_B T  \ \gg \  \max_i \Delta_i\, .
\end{eqnarray}
In terms of scaling variables, this condition may be written as
$|u_i| \ll T^{y_i/z}$ for all relevant  $i$.
Thus, the quantum-critical region occupies a high-dimensional wedge in the enlarged parameter space spanned by $\{u_1,\dots,u_N,T\}$, which collapses onto the critical point as $T \to 0$.
	
If one of the destabilizing directions is marginally relevant rather than strictly relevant, the above power-law scaling form must be modified. For a marginally relevant coupling $g_m$ obeying
\begin{eqnarray}
	\frac{d g_m}{d\ell}
	\ =\ a g_m^3\, , \qquad a \ > \ 0\,  ,
\end{eqnarray}
one obtains $g_m(\ell)^{-2}= g_{m,0}^{-2} -2a\ell$.
At finite temperature, the RG flow is cut off at
\begin{eqnarray}
	\ell_T \ \sim \ \frac{1}{z}\ln\frac{\Lambda}{T}\, ,
\end{eqnarray}
so observables depend on the running coupling $g_m(\ell_T)$ rather than on a simple scaling combination $g_m/T^{y_m/z}$. Such a marginally relevant direction can produce a long pseudo-critical regime: the flow may remain close to the fixed point over a large but finite RG time before ultimately escaping toward strong coupling or another infrared basin of attraction.
	
Experimentally, a source-like QCP is maximally fine-tuned. A conventional one-parameter QCP can be reached by tuning a single relevant scaling field to zero. By contrast, a total-repeller QCP with $N$ relevant directions requires the simultaneous tuning of all $N$ independent scaling fields. If an experiment varies only one control parameter, such as doping, pressure, or magnetic field, the resulting trajectory is a one-dimensional path through the higher-dimensional space of scaling fields. Unless this path passes through the QCP, the apparent quantum-critical regime terminates at a finite crossover temperature $T^\ast$, below which the RG flow is diverted toward one of the stable infrared phases.
	
Source-like QCPs may therefore appear naturally at multicritical junctions, where several phase boundaries meet and more than one physical tuning parameter is required to access the critical point. For example, a point repulsive in two independent directions can occur near a bicritical or tetracritical structure involving two competing ordered phases, such as superconductivity and antiferromagnetism, together with a disordered phase. In such cases, the observed fan-like regime differs from the ordinary quantum-critical fan of a codimension-one saddle: it is governed by the finite-temperature neighborhood of a high-codimension source-like fixed point, so access to the scaling regime requires simultaneous suppression of several relevant perturbations.

\section{$\omega/T$ Scaling Near a Total-Repeller QCP}
\label{sec:omega-over-T}
	
The phenomenon of $\omega/T$ scaling reflects finite-temperature scaling near a scale-invariant fixed point when temperature provides the dominant infrared cutoff. In such a regime, the frequency and temperature dependence of a dynamical response function enter through the ratio $\hbar\omega/k_B T$, up to an overall power of temperature fixed by the scaling dimension of the operator being probed. For the schematic case of two relevant scaling fields, denoted by $u_1$ and $u_2$, with RG eigenvalues $y_1>0$ and $y_2>0$, the dynamical susceptibility obeys the generalized scaling form
\begin{eqnarray}
	\chi(\omega,T,u_1,u_2)
	\ =\  T^{-\alpha} \mathcal F \left(
	\frac{\hbar\omega}{k_B T}, \frac{u_1}{T^{y_1/z}},
	\frac{u_2}{T^{y_2/z}} \right),
\end{eqnarray}
where $\alpha$ is the susceptibility exponent associated with the scaling dimension of the corresponding operator, and $z$ denotes the dynamical critical exponent. The variables $u_i$ are scaling fields, rather than arbitrary microscopic couplings. Thus, both relevant directions enter as independent scaling variables and can spoil a pure $\omega/T$ collapse.
Exact $\omega/T$ scaling is recovered only when the detuning variables are small on the thermal scale,
\begin{eqnarray}
	\left|\frac{u_i}{T^{y_i/z}}\right| \ \ll \  1\, ,
	\qquad i \ = \ 1,2\,  .
\end{eqnarray}
Equivalently, the temperature must exceed the crossover scales generated by the relevant perturbations, $k_B T \gg \Delta_i$, where $\Delta_i \sim |u_i|^{z/y_i}$. In this quantum-critical regime, the scaling function reduces approximately to its critical form
\begin{eqnarray}
	\chi(\omega,T)
	\ \simeq\
	T^{-\alpha} \mathcal F_c
	\left( \frac{\hbar\omega}{k_B T} \right),
\end{eqnarray}
with corrections controlled by $u_1/T^{y_1/z}$ and $u_2/T^{y_2/z}$.
	
Because a total-repeller QCP is a source of the infrared RG flow, any nonzero relevant scaling field grows under coarse-graining. Therefore, as the temperature is lowered, even small detunings eventually become important. Once $k_B T$ falls below one of the scales $\Delta_i$, the corresponding scaling variable becomes large, the response crosses over away from the critical form, and the $\omega/T$ collapse fails. For a more standard codimension-one QCP,
a single relevant scaling field must be tuned to zero, while irrelevant perturbations decay under the RG flow.
	
By contrast, a total-repeller QCP requires the simultaneous suppression of all relevant scaling fields. The quantum-critical regime therefore forms a high-dimensional wedge in $(u_1,u_2,T)$ space, rather than the robust codimension-one fan associated with an ordinary QCP. An experiment that varies only one control parameter samples a one-dimensional path through this higher-dimensional space. Unless this path passes sufficiently close to the fixed point in all relevant directions, the apparent scaling regime is cut off at a finite crossover temperature $T^\ast$.
	
Consequently, pure $\omega/T$ scaling near a total-repeller QCP is not structurally stable under generic perturbations. It can occur on the exactly tuned critical trajectory, and it may also appear over an intermediate quantum-critical window when the detunings are small compared with the thermal scale. However, it is not generically robust as $T\to0$, because the relevant perturbations ultimately drive the system away from the fixed-point neighborhood.
	
If one of the destabilizing directions is marginally relevant rather than strictly relevant, the power-law scaling variable $u_i/T^{y_i/z}$ must be replaced by the corresponding running coupling evaluated at the thermal RG scale. For a marginally relevant coupling $g_m$ satisfying
\begin{eqnarray}
	\frac{d g_m}{d\ell}
	\ =\
	a g_m^3\, , \qquad a \ > \ 0\, ,
\end{eqnarray}
one obtains
\begin{eqnarray}
	\frac{1}{g_m(\ell)^2}
	\ =\
	\frac{1}{g_{m,0}^2} \ -\ 2a\ell .
\end{eqnarray}
At finite temperature, the flow is cut off at
\begin{eqnarray}
	\ell_T
	\ \sim\
	\frac{1}{z} \ln\frac{\Lambda}{k_B T}\, .
\end{eqnarray}
So the susceptibility takes the schematic form
\begin{eqnarray}
	\chi(\omega,T,u,g_m)
	\ =\
	T^{-\alpha} \mathcal F \left( \frac{\hbar\omega}{k_B T},
	\frac{u}{T^{y/z}}, g_m(\ell_T) \right).
\end{eqnarray}
This produces an extended but finite pseudo-critical regime. The flow remains close to the fixed point until the marginal coupling becomes of order unity, at the scale
\begin{eqnarray}
	T^\ast
	\ \sim\
	\Lambda \exp\left[ -\frac{z}{2a g_{m,0}^2} \right] .
\end{eqnarray}
Thus, marginal relevance can generate an apparent $\omega/T$-scaling window, but the scaling is ultimately cut off at sufficiently low energies.
	
We are therefore led to conclude that, although a total-repeller QCP can in principle exhibit $\omega/T$ scaling, such behavior is not intrinsically robust. Generically, it appears only as an intermediate-scale or pseudo-critical phenomenon that is progressively suppressed as the relevant or marginally relevant perturbations grow under the RG flow. This conclusion contrasts with the proposal of Ref.~\cite{YY-Chang.25}, according to which $\omega/T$ scaling and the concomitant ``Planckian dissipation'' emerge robustly in the vicinity of the total-repeller QCP $C_r$.
	
{We emphasize, however, that the above argument should be interpreted as a statement about the generic scaling structure in the vicinity of a total-repeller fixed point, rather than as a general prohibition of every possible manifestation of apparent $\omega/T$ scaling. In principle, a particular observable may become independent of one or more relevant scaling fields if its dependence on the corresponding running couplings cancels out. In that case, the observable could retain the scaling form
\begin{eqnarray}
	\chi(\omega,T)
	\ \sim\
	T^{-\alpha} G\left(\frac{\hbar\omega}{k_B T}\right)\, ,
\end{eqnarray}
even though the full RG trajectory is flowing away from the fixed point. Such a cancellation would, however, be specific to the observable under consideration and would not imply that the total-repeller QCP governs the asymptotic low-temperature physics. A contribution that is independent of both temperature and frequency can, of course, always be written in the above scaling form with $\alpha=0$ and constant $G$, but this is not what is meant by dynamical $\omega/T$ scaling. Rather, the latter refers to a nontrivial dependence on $\omega/T$, or to a nontrivial temperature prefactor, reflecting the dynamics of a scale-invariant regime. Other observables, or the same observable once subleading corrections are retained, will generically remain sensitive to the crossover scales generated by the relevant perturbations.

It has recently been argued that such an observable-specific cancellation may indeed occur. In particular, Ref.~\cite{YY-Chang.25} proposes that the transport scattering rate becomes independent of the coupling $g$, so that all bare couplings flowing toward the U(1) FL$^*$ fixed point exhibit the same Planckian strange-metal behavior. If correct, this mechanism would evade the generic breakdown of $\omega/T$ scaling expected near a total-repeller QCP~\cite{private_Chung}. We briefly discuss this mechanism in Appendix~\ref{sec:appD}.

Thus, in the absence of either an explicit observable-specific cancellation mechanism or a distinct strong-coupling fixed point that renders the observable insensitive to the relevant perturbations, $\omega/T$ scaling near a total-repeller QCP is not structurally stable in the limit $T\to0$. Such scaling may be observed along the precisely tuned critical trajectory, or over an extended intermediate-temperature regime, but the generic expectation is that relevant or marginally relevant perturbations ultimately destroy the scaling collapse.}

To understand the origin of this discrepancy, we now turn to the RG flow equations of Ref.~\cite{YY-Chang.25}.	

\section{The RG equations and their solution
	\label{sec:solution}}
	
We now consider the RG flow equations associated with the $t$-$J$ model,  which take the form
	\begin{eqnarray}
		\frac{dg_1}{d\ell} \ &=& \ -\left(\frac{\epsilon}{2}\right)g_1 \ + \ g_1^3\ = \  \beta_1(g_1,g_2)\, ,\label{eq:flowEquations} \\[2mm]
		\frac{dg_2}{d\ell} \ &=& \ -\left(\frac{\epsilon}{2}\right)g_2 \ + \ g_1^2 g_2 \ + \ \frac{3}{2}g_2^3 \ = \
		\beta_2(g_1,g_2) \, ,
		\label{eq:A2}
	\end{eqnarray}
	where $\epsilon=d-z=1$. {The system of Eqs.~\eqref{eq:flowEquations}--\eqref{eq:A2} constitutes one of the central and technically significant results of Ref.~\cite{YY-Chang.25}.}
	
	The ensuing fixed points are determined by the nodes of the beta functions:
	\begin{eqnarray}
		\label{eq:A3}
		\left.\frac{dg_1}{d\ell}\right|_{g_j^*}&=0\, ,\\
		\left. \frac{dg_2}{d\ell}\right|_{g_j^*}&=0\, .
		\label{eq:A4}
	\end{eqnarray}
	In particular, Eq.~\eqref{eq:A3} results in $g_1^{*1}=0$ and $g_1^{*2,3}=\pm \sqrt{\epsilon/2}$ while Eq.~\eqref{eq:A4} leads to  $g_2^{*1}=0$ and $g_2^{*2,3}=\pm \sqrt{2/3}\sqrt{\epsilon/2-{g_1^*}^2}$. Two of the fixed points occur at negative coupling, either $g_1$ or $g_2$.
	In these equations, $g_1$ corresponds to $\bar{g}\rho_0$ and $g_2$ to $J \rho_0$.
	%
	\subsection{Stability analysis of fixed points}
	
	We will refer to the finite fixed points identified above as \textit{Gaussian}, \textit{isotropic}, and \textit{Ising-like} fixed points. In the next section, we will provide a justification for this terminology, and a discussion of potential fixed points at infinite coupling strengths can be found in subsection \ref{subsec:compact}.
	
	The fixed point at $\mathbf{g}^* = (g_1^{*1}, g_2^{*1}) = (0, 0) = G$ will be called the \textit{Gaussian} fixed point. The \textit{isotropic} interacting fixed points are denoted as $H_\pm = \left(\pm\sqrt{\frac{\epsilon}{2}}, 0\right)$, while we will refer to the fixed points $I_\pm = \left(0, \pm\sqrt{\frac{\epsilon}{3}}\right)$ as \textit{Ising-like}.
	In the vicinity of a  fixed point $\mathbf{g}^*$, we have
	\begin{eqnarray}
		\frac{d}{d\ell} \ \!\delta \mathbf{g} \ = \ M(\mathbf{g}^*) \delta \mathbf{g}\, ,
	\end{eqnarray}
	where $\delta \mathbf{g}=\mathbf{g}-\mathbf{g}^*$. Thus, the Jacobian matrix $M(\mathbf{g}^*)$ governs the local flow. One finds
	\begin{eqnarray}
		M(g_1,g_2) \ = \
		\begin{pmatrix}
			\frac{\partial \beta_{g_1}}{\partial g_2} & \frac{\partial \beta_{g_1}}{\partial g_2}
			\\[2mm]
			\frac{\partial \beta_{g_2}}{\partial g_1} & \frac{\partial \beta_{g_2}}{\partial g_2}
		\end{pmatrix} \ = \
		\begin{pmatrix}
			-\frac{\epsilon}{2}+3g_1^2 & 0
			\\[2mm]
			2g_1 g_2 & -\frac{\epsilon}{2}+g_1^2+\frac{9}{2} g_2^2
		\end{pmatrix} .
	\end{eqnarray}
	At the Gaussian fixed point
	\begin{eqnarray}
		M_G \ = \
		\begin{pmatrix}
			-\frac{\epsilon}{2} & 0
			\\[2mm]
			0 & -\frac{\epsilon}{2}
		\end{pmatrix}.
	\end{eqnarray}
	Thus, the Gaussian fixed point is locally stable (for increasing  $\ell$).
	At the isotropic fixed point
	\begin{eqnarray}
		M_H \ = \
		\begin{pmatrix}
			\epsilon & 0
			\\[2mm]
			0 & 0
		\end{pmatrix},
	\end{eqnarray}
	and the $g_2$ direction is marginal at linear order. Expanding the flow near $H_\pm$ gives
	\begin{eqnarray}
		\frac{dg_2}{d\ell} \ \simeq \ \frac{3}{2} g_2^3\, .
	\end{eqnarray}
	Thus, the  anisotropy is marginally relevant and the isotropic interacting fixed point is unstable.
	At the Ising-like fixed points
	\begin{eqnarray}
		M_I \ = \
		\begin{pmatrix}
			-\frac{\epsilon}{2} & 0
			\\[2mm]
			0 & \epsilon
		\end{pmatrix}.
	\end{eqnarray}
	These are saddle points with one relevant and one irrelevant direction.
	Consequently, the only stable finite-coupling fixed point is Gaussian.
	
	\subsubsection{Physical interpretation of relevant directions}
	%
	The RG equations describe only the interaction sector of the theory. A complete quantum-critical description would, in general, include an additional control parameter $r$ that tunes the system through the transition
	\begin{eqnarray}
		\frac{dr}{d\ell} \ = \ y_r r \ + \  \cdots\,  .
	\end{eqnarray}
	Consequently, the number of relevant directions associated with the interaction couplings ($g_1,g_2$)
	should not be confused with the thermal or non-thermal scaling field that drives the phase transition itself. Our analysis focuses exclusively on the stability properties of the interaction sector. The key observation is therefore not that $H_{\pm}$ possesses a relevant direction, but rather that they exhibit an additional marginally relevant instability within the interaction sector.

	\subsection{Analytical solution of the RG equations}
	%
	Physically, the exchange constant in the $t$-$J$ model is antiferromagnetic ($J>0$). In the representation employed in Ref.~\cite{YY-Chang.25}, this choice ensures convergence of the functional integral without the need for contour deformations or the introduction of imaginary Hubbard--Stratonovich fields.  No such restriction applies to $\bar{g}$.
	In what follows, we restrict attention primarily to non-negative couplings, $g_{i}\geq 0,~(i=1,2)$.
	
	To analyze the structure of the fixed points, we solve the coupled Eqs.~(\ref{eq:flowEquations}) and (\ref{eq:A2}). We begin with Eq.~(\ref{eq:flowEquations}), which after separation of variables can be rewritten as
	\begin{eqnarray}
		d\ell \ &=& \ \frac{dg_1}{g_1(g_1^2 - {\epsilon}/{2})} \ = \  -dg_1\frac{ 2}{\epsilon}\left[\frac{1}{g_1} \ - \ \frac{g_1}{g_1^2- {\epsilon}/{2}} \right]\, .
	\end{eqnarray}
	Here, the manipulations are justified only for $g_1 \neq \sqrt{\epsilon/2}$.
	By integrating both sides, we get
	\begin{eqnarray}
		\log \frac{g_1}{\sqrt{|g_1^2 - \epsilon/2|}} \ = \ - \frac{\epsilon}{2} \ \!\ell \ + \ C\, ,
	\end{eqnarray}
	where $C$ is an integration constant. The latter can equivalently be rewritten as
	\begin{eqnarray}
		\frac{g_1^2}{|g_1^2 - \epsilon/2|} \ = \ K^2 e^{-\epsilon \ell }\, ,
		\label{A.7.uk}
	\end{eqnarray}
	with $K = e^{C}$. Now, we can distinguish two cases: $g_1 > \sqrt{\epsilon/2}$ and $g_1 < \sqrt{\epsilon/2}$. For the case   $g_1 > \sqrt{\epsilon/2}$, the implicit Eq.~(\ref{A.7.uk}) can be resolved in the explicit form
	\begin{eqnarray}
		\frac{g_1^2}{g_1^2 - \epsilon/2} \ = \ K^2 e^{-\epsilon \ell } \quad \quad \Leftrightarrow \quad \quad  g_1^2 \ = \ \frac{\epsilon K^2 }{2(K^2  - e^{\ \! \epsilon \ell })} \, .
		\label{A.8.kl}
	\end{eqnarray}
	Similarly, for $g_1 < \sqrt{\epsilon/2}$, we have
	\begin{eqnarray}
		\frac{g_1^2}{\epsilon/2 - g_1^2 } \ = \ K^2 e^{-\epsilon \ell } \quad \quad \Leftrightarrow \quad \quad  g_1^2 \ = \ \frac{\epsilon K^2 }{2(K^2  + e^{\ \! \epsilon \ell })} \, .
		\label{A.9.kl}
	\end{eqnarray}
	Note that when $\ell \rightarrow -\infty$ (\textit{i.e.}, the starting point of the RG flow), $g_1$ from Eq.~(\ref{A.8.kl}) tends to $(\sqrt{\epsilon/2})_+$, while  $g_1$ from Eq.~(\ref{A.9.kl}) tends to $(\sqrt{\epsilon/2})_-$ (irrespective of $K$). So the RG fixed point with $g_1 = \sqrt{\epsilon/2}$ is a repulsive fixed point.
	
	On the other hand, when $\ell \rightarrow \infty$ (i.e. the end point of the RG flow) $g_1$ from Eq.~(\ref{A.9.kl}) tends to $0_+$, while  $g_1$ from Eq.~(\ref{A.8.kl}) reaches an infinite value already at the finite $\ell = (\log K^2)/\epsilon$, meaning that there is no fixed point at $g_1 \rightarrow \infty$.
	
	Let us now solve the second RG equation, \textit{i.e.}, Eq.~(\ref{eq:A2}). First, we lower the order of the non-linear term by taking the substitution $u = 1/g_2$. Since we are interested only in $g_2 > 0$ or $g_2\rightarrow 0_+$, $u$ should be considered positive (continuity in the limit $g_2\rightarrow 0_+$ should be checked at the end).  With this, Eq.~(\ref{eq:A2}) can be rewritten  as
	\begin{eqnarray}
		u\frac{d u}{d \ell } \ = \ \frac{\epsilon}{2} u^2 \ - \  g_1^2 u^2 \ - \  \frac{3}{2}\, ,
		\label{A.10.cv}
	\end{eqnarray}
	which after setting $y(\ell) = u^2(\ell)$, yields the linear equation
	\begin{eqnarray}
		\frac{d y}{d \ell } \ = \ {\epsilon} y \ - \  2 g_1^2 y \ - \  {3} \ = \ -py \ - \ 3\, ,
	\end{eqnarray}
	where $p(\ell) \equiv 2 g_1^2(\ell) - \epsilon$. By defining
	\begin{eqnarray}
		a(\ell) \ = \ \exp\left(\int p(\ell) d\ell\right) \ = \  \exp(-\epsilon \ell ) \exp\left(2\int g_1^2(\ell) d\ell \right) ,
	\end{eqnarray}
	the method of variation of parameters allows to seek the solution in the form
	\begin{eqnarray}
		y(\ell)\ = \ \frac{-3 \int a(\ell) d\ell \ + \ C}{a(\ell)}\, ,
	\end{eqnarray}
	($C$ is an integration constant), or equivalently
	\begin{eqnarray}
		g_2^2 \ = \ \frac{a(\ell)}{-3 \int a(\ell) d\ell \ + \ C}\, .
	\end{eqnarray}
	In particular, for $g_1$ from Eqs.~(\ref{A.8.kl}) and (\ref{A.9.kl}), we obtain
	\begin{eqnarray}
		a(\ell) \ = \  \frac{1}{\epsilon K^2(K^2 - e^{\epsilon \ell})}     \;\;\;\; \mbox{and} \;\;\; a(\ell) \ = \ \frac{1}{\epsilon K^2(K^2 + e^{\epsilon \ell})}\, ,
	\end{eqnarray}
	respectively. With these results we get for $g_1 > \sqrt{\epsilon/2}$ and $g_1 < \sqrt{\epsilon/2}$
	\begin{eqnarray}
		g_2^2 \ = \ \frac{\epsilon K^2}{(K^2 - e^{\epsilon \ell})\left[CK^2 \epsilon \ + \ 3 \log(K^2 e^{-\epsilon \ell}- 1 )\right]}\, ,
		\label{A.16.bf}
	\end{eqnarray}
	and
	\begin{eqnarray}
		g_2^2 \ = \ \frac{\epsilon K^2}{(K^2 + e^{\epsilon \ell})\left[CK^2 \epsilon \ + \ 3 \log(K^2 e^{-\epsilon \ell}+ 1 )\right]}\, ,
		\label{A.17.bf}
	\end{eqnarray}
	respectively.
	
	To complete our analysis, we should verify that the limit $g_2\rightarrow 0_+$ is continuous. This is directly seen from Eqs.~(\ref{A.16.bf}) and~(\ref{A.17.bf}). First, Eq.~(\ref{A.17.bf}) implies that $g_2$ approaches $0_+$ continuously as $\ell \to \infty$. This corresponds to the stable Gaussian fixed point $\mathbf{g}^* = (g_1^{*1},g_2^{*1}) = (0,0)$.  Second, Eqs.~(\ref{A.16.bf}) and~(\ref{A.17.bf}) together show that $g_2$ also approaches $0_+$ continuously as $\ell \to -\infty$. This corresponds to unstable fixed point $\mathbf{g}^* = (g_1^{*2},g_2^{*1}) =(\sqrt{\epsilon/2},0)$. In particular, we conclude that in this model there exists no stable fixed point other than the trivial Gaussian one at $(0,0)$. The isotropic interacting fixed point exists only on a finely tuned manifold and is destabilized by arbitrarily weak anisotropy.

	\subsection{Numerical Solution and Poincar\'{e} Compactification}
	\label{subsec:compact}
	
\begin{figure}[ht]
		\begin{center}
			\includegraphics[scale=0.55]{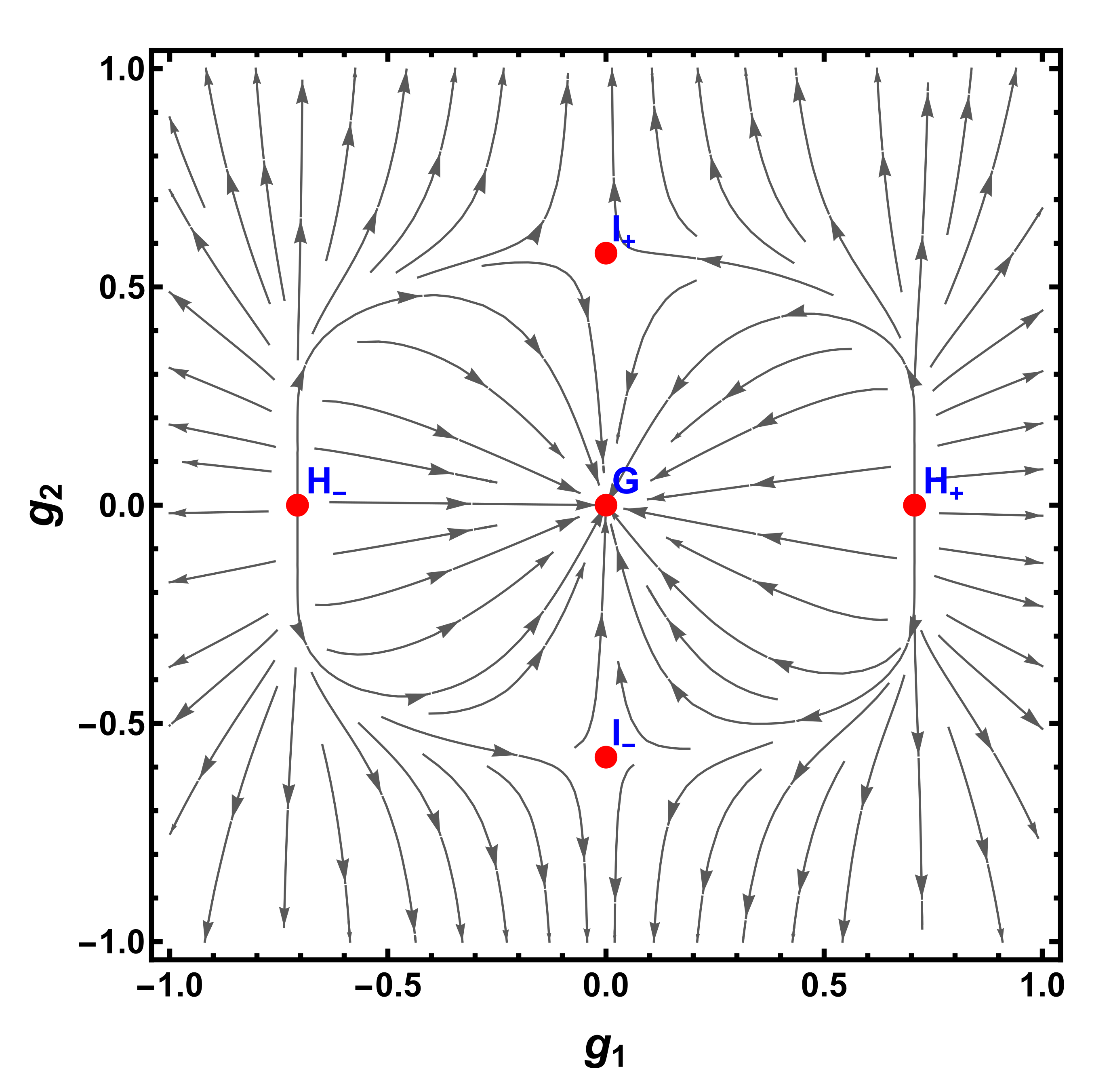}~~
			\includegraphics[scale=0.55]{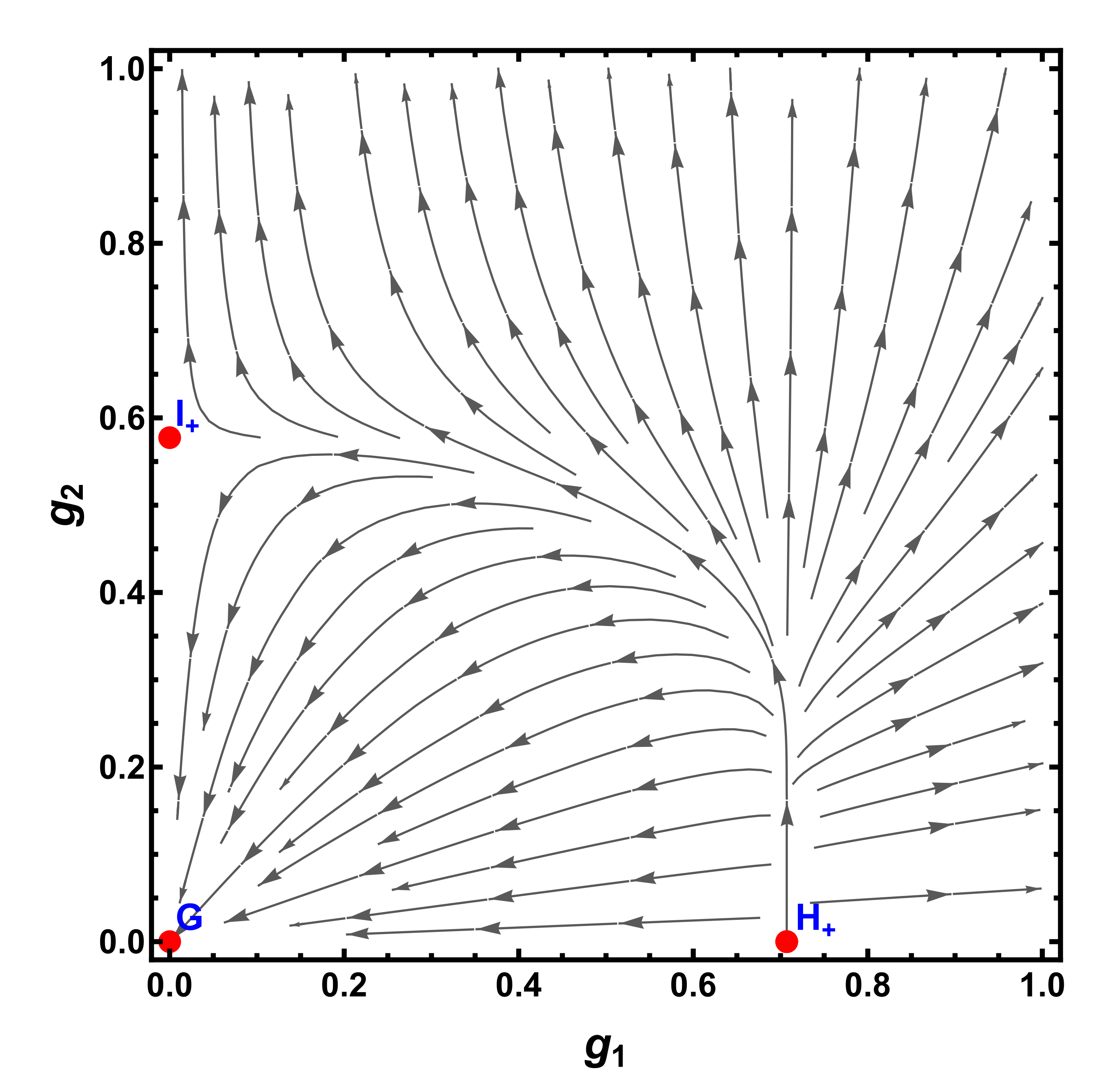}~~~
		\end{center}
		\caption{\footnotesize{Numerical phase portrait of the RG Eqs.~(\ref{eq:flowEquations}) for $\epsilon=1$. Red symbols denote finite-coupling fixed points. The Gaussian fixed point $G=(0,0)$ is the only infrared-attractive finite fixed point. The interacting fixed points $H_\pm=(\pm \epsilon/2,0)$ lie on the invariant manifold $g_2=0$ and are unstable to transverse perturbations, while the Ising-like fixed points $I_{\pm}=(0,\pm \epsilon/3)$ are saddles. The left panel shows the full range $-1 \leq g_{1,2}\leq 1$; the right panel highlights the physically relevant sector $g_1 \geq 0$, $g_2 \geq 0$.}}
		\label{fig:flow_solved}
	\end{figure}

	The RG Eqs.~(\ref{eq:flowEquations})--(\ref{eq:A2}) may also be solved numerically. Following Ref.~\cite{YY-Chang.25}, we set $\epsilon=1$. The resulting flow diagram is shown in Fig.~\ref{fig:flow_solved}, both for the domain $-1\leq g_1\leq 1$ and $-1\leq g_2\leq 1$ (left panel), and for the first quadrant $0\leq g_1\leq 1$ and $0\leq g_2\leq 1$ (right panel). The red dots indicate the fixed points identified above. We observe that there is {\em only one} attractive fixed point, namely the Gaussian fixed point  $G = (g_1^{*1},g_2^{*1}) = (0,0)$. This conclusion is fully consistent with the analytical solution obtained in the previous section.
	
	Comparing Figs.~\ref{fig:flow_solved} and~\ref{fig:RGflow}, we observe that our flow diagram differs from that reported in Ref.~\cite{YY-Chang.25}, both in the number of fixed points and in its overall topology. There is, however, agreement regarding the location and stability of the Gaussian fixed point. Focusing on the sector $g_1\geq 0$ and $g_2\geq 0$
	further suggests that the fixed point $Q$ in Fig.~\ref{fig:RGflow} should be identified with our Ising-like fixed point $I_+$.  By contrast, whereas the fixed point $P$ of Ref.~\cite{YY-Chang.25} possesses one relevant ($g_1$) and one irrelevant ($g_2$) direction,  we find that  the isotropic fixed point $H_+$ acts like a total repeller fixed point, with one relevant and one marginally relevant direction.

	In order to understand the behavior at infinite coupling, where Ref.~\cite{YY-Chang.25} identified the fixed points PSG, FL and SC, we analyze the global structure of Eqs.~(\ref{eq:flowEquations})--(\ref{eq:A2}).
	To this end, we employ the Poincar\'{e} compactification, which is a geometric tool that maps the RG flow from the infinite state space, { i.e.}, the space of couplings, onto a compact manifold. Typically, this manifold is represented by a unit hemisphere or sphere, known as the \textit{Poincar\'{e} sphere}, with the boundary at infinity mapped onto its equator. A fixed point on the equator represents an asymptotic direction of the RG flow at infinite coupling. Such a point may correspond to a candidate strong-coupling scaling regime only if the theory admits a well-defined continuation there. Otherwise, it signals a breakdown of the finite-coupling effective description.
	
\begin{figure}[h]
		\begin{center}
			\includegraphics[scale=0.6]{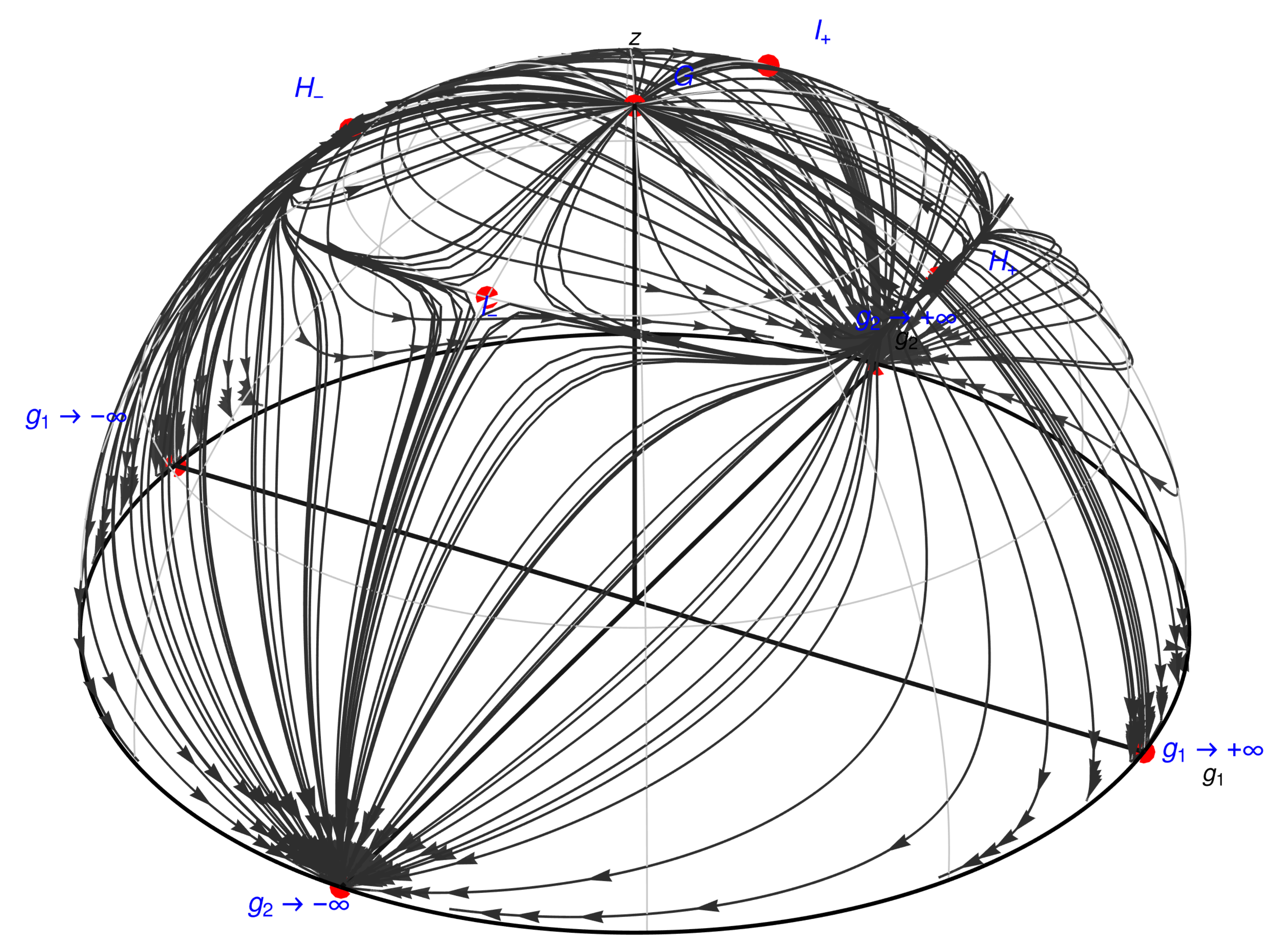}
		\end{center}
		\caption{\footnotesize{Poincar\'e compactification of the RG flow of Eqs.~(\ref{eq:flowEquations})--(\ref{eq:A2}) for $\epsilon=1$. The infinite coupling plane is mapped onto a semi-sphere with the equator representing asymptotic strong-coupling directions.
				Generic trajectories are attracted toward the $g_2$-dominated directions at infinity, while the $g_1$-axis directions act as angular saddles. The compactified flow demonstrates the absence of a stable interacting infrared fixed point and reveals the global runaway topology of the RG equations.}}
		\label{fig:compactified_a}
	\end{figure}

\begin{figure}[h]
		\begin{center}
			\includegraphics[scale=0.8]{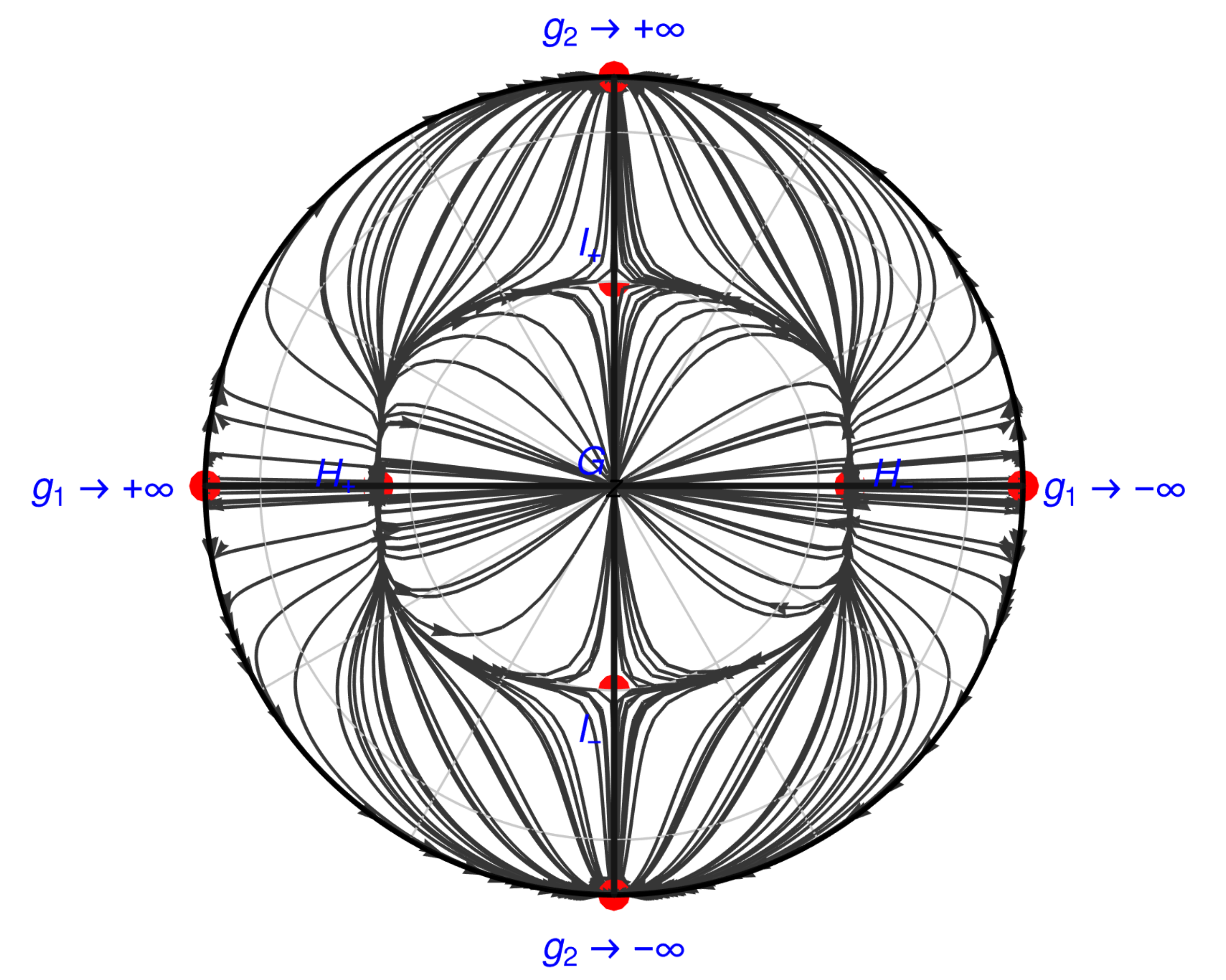}
		\end{center}
		\caption{\footnotesize{Poincar\'e compactification of the RG flow of Eqs.~(\ref{eq:flowEquations})--(\ref{eq:A2}) for $\epsilon=1$.  When viewed from below the angular flow at infinity becomes transparent.
				Generic trajectories are attracted toward the $g_2$-dominated directions at infinity, while the $g_1$-axis directions act as angular saddles. The compactified flow demonstrates the absence of a stable interacting infrared fixed point and reveals the global runaway topology of the RG equations.}}
		\label{fig:compactified}
	\end{figure}

	Let $(x,y,z)$ denote coordinates on the unit sphere. The transformation from the coupling constants $(g_1,g_2)$ to the sphere is defined as (see, e.g.~\cite{Perko2001})
	\begin{eqnarray}
		x \ = \ \frac{g_1}{\sqrt{1 + g_1^2 + g_2^2}}\, , \quad y \ = \ \frac{g_2}{\sqrt{1 + g_1^2 + g_2^2}}\, , \quad z \ = \ \frac{1}{\sqrt{1 + g_1^2 + g_2^2}}\, .
	\end{eqnarray}
	or equivalently
	\begin{eqnarray}
		g_1 \ =\ \frac{x}{z}\, ,\quad\;\;
		\; g_2 \ =\ \frac{y}{z}\, , \quad \;\;\; x^2\ + \ y^2 \ + \ z^2 \ =\ 1,
	\end{eqnarray}
	with $z>0$. To obtain a vector field that extends smoothly to the equator $z=0$, one must additionally perform a rescaling of the RG time parameter. For the flow equations~\eqref{eq:flowEquations}--\eqref{eq:A2}, the appropriate positive time-rescaling rescaling factor is
	\begin{eqnarray}
		z^2\ = \ \frac{1}{1+g_1^2+g_2^2}\, .
	\end{eqnarray}
	Under this transformation, every finite point in coupling space is mapped to the open upper hemisphere $z>0$. In the limit $\sqrt{g_1^2+g_2^2}\rightarrow\infty$, one has $z\rightarrow 0$, so that the entire ``boundary at infinity'' is represented by the equator of the sphere. The compactified RG flow is then given by
	\begin{eqnarray}
		\dot{x}
		\ &=&  \ z^3\left[
		\beta_1 \ - \ x\left(x\beta_1 \ + \ y\beta_2\right)
		\right]\, , \nonumber  \\[2mm]
		\dot{y}
		\  &=& \ z^3\left[
		\beta_2 \ - \  y\left(x\beta_1 \ + \ y\beta_2\right)
		\right]\, , \nonumber \\[2mm]
		\dot{z}
		\ &=& \  -z^4\left(x\beta_1 \ + \ y\beta_2\right)\, ,
	\end{eqnarray}
	where $\beta_i$ are evaluated at
	\begin{eqnarray}
		g_1 \ = \ \frac{x}{z}\, , \qquad g_2 \ = \ \frac{y}{z}\, ,
	\end{eqnarray}
	and the dot denotes differentiation with respect to the rescaled RG time.
	This compactification is shown in Figs.~\ref{fig:compactified_a} and~\ref{fig:compactified} for $\epsilon=1$.
	%
	The angular flow along the boundary at infinity can be obtained from writing
	\begin{eqnarray}
		g_1 \ = \ r\cos\theta\, ,
		\qquad
		g_2\ = \ r\sin\theta\, ,
	\end{eqnarray}
	and rescaling
	\begin{eqnarray}
		\frac{d \theta}{d \tau} \ = \ \frac{1}{r^2}\frac{d \theta}{d\ell}\, ,
	\end{eqnarray}
	Retaining the dominant cubic terms,
	\begin{eqnarray}
		\dot{g}_1 \ \sim \  g_1^3\, , \;\;\;\;\;
		\dot{g}_2 \ \sim \  g_1^2 g_2 \ + \  \frac{3}{2} g_2^3\, ,
	\end{eqnarray}
	one obtains the angular flow at infinity
	\begin{eqnarray}
		\frac{d\theta}{d\tau}
		\ = \
		\frac{3}{2} \cos\theta\sin^3\theta\, .
	\end{eqnarray}
	The compactified boundary contains fixed points at
	\begin{eqnarray}
		\theta \ = \ 0,\, \pi,\, \frac{\pi}{2},\, \frac{3\pi}{2}\,.
	\end{eqnarray}
	These correspond respectively to
	\begin{eqnarray}
		g_1 \ = \ \pm\infty\, ,
		\qquad
		g_2 \ = \ \pm\infty\, .
	\end{eqnarray}
	The $g_2=\pm\infty$ directions are stable sinks on the compactified boundary, while the $g_1=\pm\infty$ directions are unstable angular saddles. Generic trajectories therefore flow toward $g_2$-dominated infinity. The compactified flow is shown in Figs.~\ref{fig:compactified_a} and~\ref{fig:compactified}.
	%
%
The analytical and numerical solutions of the RG equations of Ref.~\cite{YY-Chang.25}, Eqs.~(\ref{eq:flowEquations}) and~(\ref{eq:A2}), lead to an RG flow diagram that differs substantially from that reported in Ref.~\cite{YY-Chang.25}, both with regard to the fixed-point structure at finite coupling strength and, as revealed by the Poincar\'{e} compactification, the asymptotic behavior of the flow at infinity.
The compactified flow does not reveal distinct stable equatorial fixed points corresponding to PSG, FL, and SC. Instead, the generic strong-coupling fate is a runaway trajectory toward the anisotropy-dominated sector at infinity.
To understand the origin of this discrepancy, we have analyzed the transformation properties of RG flow equations under field redefinitions.
In  Appendices~A and~B, we argue that the number and stability properties of fixed points, {i.e.}, whether a given coupling corresponds to a relevant, irrelevant, or marginally relevant direction, are invariant under changes of renormalization scheme, provided that the transformations preserve the physical content of the theory.

\section{A quantum analogue of anisotropy-driven runaway \label{runaway}}
	
	In this section, we argue  that the RG Eqs.~(\ref{eq:flowEquations})--(\ref{eq:A2}) describe a symmetry-breaking process associated with the stability of the $O(n)$-symmetric fixed point under cubic perturbations, as pioneered by Aharony~\cite{Bruce.75,Aharony.73}.

\subsection{Mapping to the Cubic Anisotropy Problem}
	
For a system with $n$ real-valued field components, the Ginzburg--Landau--Wilson Hamiltonian in the presence of cubic anisotropy is given by
\begin{eqnarray}
	\mathcal{H} \ = \  \int d^d x \left[ \frac{1}{2}(\nabla \vec{\phi})^2 \ + \ \frac{r}{2}\sum_i \phi_i^2 \ + \ u\left(\sum_i \phi_i^2\right)^{\!2} \ + \  v \sum_i \phi_i^4 \right],
	\end{eqnarray}
where $u$ governs the isotropic $O(n)$-symmetric quartic interaction, and $v$ parametrises the leading symmetry-breaking perturbation corresponding to cubic anisotropy.

By identifying $g_1$ and $g_2$ in Eqs.~(\ref{eq:flowEquations}) and (\ref{eq:A2}) as linear combinations of the isotropic coupling $u$ and the anisotropic coupling $v$, the fixed point $(\sqrt{\epsilon/2},0)$ is seen to describe a critical isotropic sector ($u \neq 0$) with the cubic anisotropy tuned to zero, $v = 0$. The cubic term $\tfrac{3}{2}g_2^3$ represents the self-interaction of the anisotropic mode, and its coefficient controls whether the anisotropy is relevant or irrelevant along the RG flow.
	
In Aharony's analysis, the stability of the $O(n)$-symmetric fixed point is determined by the scaling exponent $y_v$ associated with the cubic anisotropy $v$. For $y_v>0$, the cubic perturbation is relevant and the $O(n)$ fixed point becomes unstable. In the system of Eq.~(\ref{eq:A2}), the eigenvalue associated with $g_2$ vanishes at the linear level, rendering the anisotropic direction marginal. Its subsequent behavior is governed by the nonlinear term proportional to $g_2^3$. This is analogous to the situation at $n=n_c$ in Aharony's analysis, where the isotropic and cubic fixed points coincide. The resulting runaway RG flow was studied by Rudnick~\cite{Rudnick.78}.

\subsection{Invariant Critical Manifold, runaway flow and marginal instability}
	
A central feature of the flow equations is the existence of an invariant manifold defined by $g_2 = 0$. Substituting this condition into the second beta function [cf Eq.~(\ref{eq:A2})], i.e
\begin{eqnarray}
	\beta_2(g_1, g_2) \ = \  g_2 \left( -\frac{\epsilon}{2} \ + \  g_1^2 \ + \ \frac{3}{2} g_2^2 \right)\, ,
\end{eqnarray}
shows that trajectories originating on the $g_2 = 0$ axis  remain confined to this manifold under the RG flow. Restricting the dynamics to this subspace, the flow reduces to a single-coupling equation
\begin{eqnarray}
	\frac{dg_1}{d\ell} \ = \  -\frac{\epsilon}{2}g_1 \ + \  g_1^3\, ,
\end{eqnarray}
which exhibits the interacting Wilson--Fisher fixed points located at $H_\pm = \left( \pm\sqrt{\epsilon/2}, 0 \right)$.
	
Within this invariant manifold, these fixed points govern scale-invariant critical behavior. However, their stability with respect to transverse perturbations ($g_2 \neq 0$) is crucial. In the vicinity of the interacting fixed point, where $g_1^2 \simeq \epsilon/2$, the anisotropy flow reduces to
\begin{eqnarray}
	\frac{dg_2}{d\ell} \ \simeq \  \frac{3}{2} g_2^3\, .
\end{eqnarray}
Consequently, the transverse perturbation is marginal at linear order, with its behavior determined by the leading nonlinear contribution. In particular, the flow is governed by a marginal cubic term.	
	
Accordingly, the RG flow approaches the vicinity of $H_\pm$ and exhibits an extended regime of pseudo-critical scaling. Ultimately, however, the flow departs from the vicinity of the fixed point due to the slow nonlinear growth of the transverse coupling.
	
Integrating this equation yields
\begin{eqnarray}
	\frac{1}{g_2(\ell)^2} \ = \  \frac{1}{g_{2,0}^2} \ - \  3\ell\, ,
\end{eqnarray}
where $g_{2,0}$ denotes the initial value of the transverse coupling at the microscopic scale. The coupling diverges at a finite RG scale $\ell_* \sim (3g_{2,0}^2)^{-1}$. The associated correlation length is therefore
\begin{eqnarray}
	\xi \ \sim \ e^{\ell_*} \ \sim \  \exp\left( \frac{1}{3g_{2,0}^2} \right)\, .
\end{eqnarray}
This exponentially large but finite correlation length is characteristic of a marginally relevant perturbation. The system exhibits an extended pseudo-critical regime over a broad intermediate range of scales, but ultimately departs from the critical manifold as the transverse coupling grows. This mechanism, where a system displays near-critical behavior before the runaway flow of a marginal operator drives it away from criticality,  is indicative of weakly first-order behavior.

\subsection{Evidence for weakly first-order behavior}
	
The runaway flow strongly suggests weakly first-order behavior.
At a genuine continuous transition, RG trajectories terminate at a stable scale-invariant infrared fixed point. Here, however, no such stable fixed point exists outside the Gaussian basin. Instead, trajectories initially approach the isotropic interacting fixed point and then slowly escape because of the marginally relevant anisotropy.
The resulting physics is pseudo-critical,
$a_{\rm UV} \ll |x| \ll \xi$ with an exponentially large but finite correlation length.
Here, $a_{\rm UV}$  is the microscopic cutoff length, such as the lattice spacing, and $|x|$ denotes the real-space observation scale, e.g., the separation at which correlation functions are measured.
For distances much larger than the microscopic scale but still much smaller than the correlation length, the system behaves approximately as if it were at a continuous critical point. Correlation functions in this regime exhibit apparent scale invariance and critical scaling.
Because the anisotropy is marginally relevant, the RG flow eventually runs away instead of ending at a stable infrared fixed point. Consequently, $\xi$ is finite and for distances
$|x|>\xi$, the pseudo-critical scaling breaks down and the true noncritical behavior emerges.
	
This mechanism is closely analogous to the fluctuation-induced first-order transitions discussed by Coleman and Weinberg~\cite{Coleman.73} and by Halperin, Lubensky, and Ma~\cite{Halperin.74}. In the Coleman--Weinberg mechanism, fluctuations destabilize the naive critical theory and generate a first-order transition through radiative corrections to the effective potential. Similarly, Halperin, Lubensky, and Ma demonstrated that gauge fluctuations destabilize the superconducting critical point, producing runaway RG flow and fluctuation-induced first-order behavior.
	
The RG Eqs.~(\ref{eq:flowEquations})--(\ref{eq:A2}) exhibit essentially the same features, {i.e.},
the existence of a nearly critical interacting fixed point, a marginal instability, slow runaway flow, an exponentially large crossover scale, and the absence of asymptotic scale invariance.
The resulting transition is thus characteristic of weakly first-order behavior.	
	
\subsection{Relation to Bruce--Aharony and Rudnick}
	
Bruce and Aharony~\cite{Bruce.75} analyzed coupled-order-parameter systems in which anisotropic perturbations compete with an isotropic critical fixed point. Their work showed that the asymptotic critical behavior depends sensitively on the stability of symmetry-breaking perturbations: the system may display isotropic criticality, cubic criticality, or, in appropriate regions of parameter space, first-order behavior associated with the failure to reach a stable critical fixed point.
	
The present RG equations exhibit an analogous instability mechanism. The subspace $g_2=0$ defines an invariant critical manifold, within which the interacting fixed points
\begin{eqnarray}
	H_\pm \ = \ \left(\pm\sqrt{\epsilon/2},0\right)\, ,
\end{eqnarray}
control an apparent scale-invariant regime. However, these fixed points are not stable against generic transverse perturbations. Near $H_\pm$, the flow of the anisotropy reduces to
\begin{eqnarray}
	\frac{dg_2}{d\ell} \ \simeq\  \frac{3}{2}g_2^3\, ,
\end{eqnarray}
so that the anisotropic direction is marginal at linear order but becomes marginally relevant at nonlinear order. As a result, trajectories that pass close to the invariant critical manifold exhibit an extended pseudo-critical regime before eventually escaping toward strong coupling.
		
This structure is closely related to the anisotropy-driven mechanism studied by Rudnick~\cite{Rudnick.78}. Rudnick considered a two-component Landau--Ginzburg--Wilson theory with cubic anisotropy and showed that sufficiently large anisotropy can remove the RG fixed point controlling the continuous transition. In that case, an explicit construction of the free energy verifies that the loss of the fixed point corresponds to a first-order transition. To leading order in the $(4-\epsilon)$-expansion, the recursion relations for the quartic couplings $u(\ell)$ and $v(\ell)$ on the critical hypersurface are
\begin{eqnarray}
	\frac{du}{d\ell} \ &=&\ \epsilon u\ - \ 36u^2 \  - \ v^2\, , \nonumber \\[2mm]
	\frac{dv}{d\ell} \ &=&\ \epsilon v \ - \ 24uv\ - \ 8v^2\, .
\end{eqnarray}
These equations are not algebraically identical to the present flow equations, nor is the marginality structure the same: in Rudnick's case the cubic perturbation at the isotropic fixed point is relevant already at linear order. Nevertheless, the physical mechanism is analogous. In both cases, an apparently critical interacting fixed point is destabilized by an anisotropic perturbation, and the flow ultimately avoids a stable scale-invariant infrared fixed point.
	
Introducing $X=g_1^2$ and $Y=g_2^2$, the present equations become
\begin{eqnarray}
	\frac{dX}{d\ell} \ &=&\ -\epsilon X \ + \ 2X^2\, , \nonumber \\[2mm]
	\frac{dY}{d\ell} \ &=&\ -\epsilon Y\ + \ 2XY \ + \ 3Y^2\,  .
\end{eqnarray}
This representation makes the analogy transparent within a fixed sign sector of the original $(g_1,g_2)$ plane. The fixed point $X=\epsilon/2,\ Y=0$ lies on the invariant isotropic manifold, while the anisotropy $Y$ is marginal at linear order and driven away by the nonlinear $3Y^2$ term. Thus, the correspondence to Rudnick should be understood at the level of RG mechanism and flow organization, rather than as an exact algebraic or global topological equivalence.
	
The Poincar\'{e} compactification further supports this interpretation by displaying the global fate of the runaway trajectories. Once the transverse anisotropy is generated, the relevant trajectories do not approach a stable finite strong-coupling fixed point. Instead, they escape toward an anisotropy-dominated sector at infinity. This absence of an asymptotically scale-invariant infrared endpoint provides strong RG evidence for pseudo-critical behavior and is naturally interpreted, by analogy with Rudnick's analysis, as indicative of a fluctuation-induced weakly first-order transition.
\subsection{Quantum-Critical Interpretation}

A significant conceptual distinction exists between the present system and the classical thermal transitions previously examined by Bruce, Aharony, and Rudnick.
In the classical case, RG analyses typically target the universal scaling of the free energy near a finite-temperature critical point.
By contrast, in the current framework, the RG equations describe the evolution of the interaction sector within a quantum Hamiltonian controlling a QCP.
Here, the physical tuning parameter, such as pressure, magnetic field, or doping, is distinct from the interaction couplings $g_1$ and $g_2$.
Consequently, the full RG structure must be understood as a coupled system:
\begin{eqnarray}
	\frac{dr}{d\ell} \ = \  y_r r \ + \  \cdots\, ,
\end{eqnarray}
where $r$ represents the non-thermal control parameter tuning the system toward the QCP, evolving alongside the flow of the interaction couplings.
This distinction importantly influences the interpretation of fixed points.
The Ising-like fixed points,
$I_\pm = (0, \pm\sqrt{\epsilon/3})$,
may represent standard QCP universality classes provided that the physical tuning parameter serves as the sole relevant scaling field, while all other interaction-sector perturbations remain irrelevant.
In contrast, the interacting fixed points $H_\pm$ possess an inherent instability due to a marginally relevant interaction-sector perturbation.
Even when the system is tuned precisely to the critical manifold ($r=0$), the RG flow eventually drives the system away from these fixed points.
Consequently, the observed scaling behavior is merely pseudo-critical rather than truly scale-invariant.
It is crucial to recognize that the weakly first-order nature of this transition does not arise simply because the fixed point possesses a relevant thermal direction, as this is a standard feature of any ordinary critical point.
Rather, the defining characteristic here is that the interaction sector itself remains unstable under RG flow.
In this sense, the present system serves as a quantum analogue to the anisotropy-driven runaway scenarios described in classical statistical mechanics.
	
The above analysis demonstrates that the RG Eqs.~(\ref{eq:flowEquations}) and (\ref{eq:A2}) constitute a minimal RG normal form that realizes the mechanism famously identified by Coleman and Weinberg, Halperin, Lubensky, and Ma, and subsequently explored by Bruce, Aharony, and Rudnick.

\section{Conclusion}
\label{Sec:conclusion}
	
In this work we have analyzed the renormalization-group structure of the dynamical charge-Kondo breakdown scenario proposed for cuprate superconductors. By studying the resulting coupled RG flow equations, {recently derived in the seminal work of Chang~\emph{et al.}~\cite{YY-Chang.25},} both locally and globally, we have determined the stability properties of the fixed points as well as the global topology of the flow.
	
Our analysis shows that the interacting finite-coupling fixed point proposed as the origin of quantum-critical metallic behavior is not asymptotically stable. Although the RG flow admits an invariant critical manifold containing interacting fixed points, perturbations transverse to this manifold generate a marginally relevant instability that ultimately drives the system away from criticality. As a consequence, the apparent critical scaling associated with these fixed points is only transient. It arises from a prolonged flow in the vicinity of the critical manifold, but does not correspond to a truly infrared-attractive fixed point, and should therefore be interpreted as pseudo-critical rather than genuinely scale invariant behavior.
	
The resulting RG structure is reminiscent of the anisotropy-driven runaway flows encountered in fluctuation-induced weakly first-order phase transitions. The flow initially approaches the vicinity of the interacting fixed point and exhibits an extended regime of nearly critical behavior, but ultimately escapes along an unstable direction toward strong coupling. The associated crossover scale is exponentially large, producing broad scaling regimes that may mimic quantum-critical behavior over experimentally accessible ranges while remaining fundamentally distinct from a true interacting quantum critical point. The global perspective provided by the compactified phase portrait further clarifies this picture. Rather than approaching a stable infrared scaling state, generic trajectories are attracted toward runaway directions at the boundary of coupling space. The absence of a stable interacting infrared fixed point suggests that the observed scaling behavior originates from proximity to a marginally unstable critical manifold rather than from an asymptotically critical quantum state.
	
The results obtained indicate that extended scaling behavior alone does not constitute evidence for a stable interacting quantum critical point. More generally, they illustrate how global RG topology can provide information that is inaccessible from a purely local fixed-point analysis. In this sense, the charge-Kondo-breakdown scenario provides an instructive example of how pseudo-critical behavior can arise from marginal instabilities and runaway RG flows in strongly correlated quantum systems, while simultaneously serving as a testing ground for the analysis of such phenomena.


\section*{Acknowledgements}
The authors are grateful to C.-H. Chung for stimulating discussions.

\paragraph*{Funding information.}
S.K. acknowledges support by the National Science and Technology Council of Taiwan through Grant No.\ 112-2112-M-A49-MY4. P.J. acknowledges support by the Czech Science Foundation Grant (GAČR), Grant No.\ 25-18105S. The successful completion of this research was made possible by the academic resources and advanced research infrastructure provided by the National Center for High-Performance Computing, National Institutes of Applied Research (NIAR), Taiwan.

\appendix
\section{Topological stability of fixed points in two-dimensional RG flows}
\label{App.C.kl}
	
In this Appendix, we review some concepts required for the analysis of topological invariants associated with two-dimensional RG flows. In particular, we briefly summarize the notions that are used in the main body of the text and discuss several characteristic topological invariants of RG flows. We show that particular care must be taken when performing transformations of the coupling variables, since only transformations preserving the relevant topological structure leave these invariants unchanged. More general transformations may alter the number of singular points, their indices, or other topological characteristics of the flow, resulting in a topologically inequivalent phase portrait and, therefore, a potentially different physical theory. Although most of these concepts are standard in the theory of dynamical systems, we include this concise review for completeness and to establish the notation used throughout the paper.

Let us start by considering RG flows as autonomous dynamical systems on a two-dimensional coupling manifold $\mathcal{M} \subset \mathbb{R}^2$, namely
\begin{eqnarray}
	\frac{d g_i}{d\ell} \ = \  \beta_i(\mathbf{g}), \qquad \mathbf{g} \ = \ (g_1,g_2)\, , \quad i \ = \ 1,2\, ,
	\label{AC.1.kl}
\end{eqnarray}
with $\beta: \mathcal{M} \to T\mathcal{M}$ a smooth vector field. Fixed points $\mathbf{g}^\ast$ are defined via the relation
\begin{eqnarray}
	\beta(\mathbf{g}^\ast) \ = \ 0\, .
\end{eqnarray}
We further define the Jacobian matrix
\begin{eqnarray}
	M_{ij}(\mathbf{g}^\ast) \ = \ \left.\frac{\partial \beta_i}{\partial g_j}\right|_{\mathbf{g}^\ast}\, .
\end{eqnarray}
With this, the linearization of~(\ref{AC.1.kl}) around $\mathbf{g}^\ast$ yields the equation
\begin{eqnarray}
	\frac{d}{d\ell}\delta \mathbf{g} \ = \  M(\mathbf{g}^\ast)\,\delta \mathbf{g}\, ,
\end{eqnarray}
where  $\delta \mathbf{g}(\ell) =  \mathbf{g}(\ell) - \mathbf{g}^\ast$ denotes the deviation from the fixed-point coupling configuration.

The key concepts characterizing RG flows in two dimensions include the spectrum of the Jacobian matrix $M(\mathbf{g}^\ast)$, the sign of $\det M$, the separatrix network $\Sigma$, and the degree and homotopy class of the vector field $\beta$. We now briefly discuss each of these in turn.
\subsection*{A1: Spectrum of $M_{ij}(\mathbf{g}^\ast)$ and index of $\mathbf{g}^*$}
	
Denoting  the eigenvalues of $M$ as $\lambda_{1,2}$, one may classify the fixed points according to the signs of $\lambda_{1,2}$ as follows:
\begin{eqnarray}
	&&\mathrm{Re}(\lambda_{1,2}) \ < \ 0 \;\;\;  \Rightarrow  \;\;\; \text{stable node (IR attractor)}\, ,\nonumber \\[2mm]
	&&\mathrm{Re}(\lambda_{1,2}) \ > \ 0 \;\;\; \Rightarrow \;\;\;  \text{unstable node (IR repeller)}\, , \nonumber \\[2mm]
	&&\mathrm{Re}(\lambda_1)\,\mathrm{Re}(\lambda_2) < 0
	\;\;\; \Rightarrow \;\;\; \text{saddle point (one relevant}\nonumber \\[2mm] &&\mbox{\hspace{46mm}}\text{and one irrelevant direction)}\, .
	\end{eqnarray}
If $\mathrm{Re}(\lambda_a)=0$ for at least one eigenvalue, the corresponding direction is {\em marginal} at linear order. In this case, the RG flow is controlled by higher-order terms in the beta functions, and the direction may be classified as {\em marginally relevant}, {\em marginally irrelevant}, or {\em exactly marginal} depending on nonlinear corrections.
	
Another important concept in the classification of fixed points in {\em topological index}.
The topological index of an isolated fixed point is defined as~\cite{GuilleminPollack1974,Milnor1965Topology}
\begin{eqnarray}
	\mathrm{ind}(\mathbf{g}^\ast)
	\ = \  \mathrm{sign}\!\left[\det M(\mathbf{g}^\ast)\right]
	\ = \  \mathrm{sign}(\lambda_1 \lambda_2)
	\ \in \ \{+1,-1\}\, ,
\end{eqnarray}
which characterizes whether the linearized flow locally preserves or reverses orientation.
	
More generally, the index can be defined as the winding number of the vector field $\beta_i(\mathbf{g})$ around the fixed point.
For a compactified coupling manifold $\widehat{\mathcal{M}}$ (e.g. $S^2$), the Poincar\'{e}--Hopf theorem~\cite{Milnor1965Topology} implies
\begin{eqnarray}
	\sum_{\mathbf{g}^\ast \,:\, \beta(\mathbf{g}^\ast)=0}
	\mathrm{ind}(\mathbf{g}^\ast)
	\;=\;
	\chi(\widehat{\mathcal{M}})\, ,
	\label{App.A.56.vb}
\end{eqnarray}
relating the sum of RG fixed-point indices to the Euler characteristic of the coupling space. This theorem states that the total ``topological charge'' of all fixed points of a vector field is fixed by the topology of the space, so that fixed points cannot be created or destroyed arbitrarily: they must appear or annihilate in a way that preserves the total index.
		
More generally, the classification is stable under smooth deformations of $\beta$ preserving hyperbolicity:
\begin{eqnarray}
	\det M(\mathbf{g}^\ast) \ \neq \  0 \;\; \Rightarrow \;\;\text{structural stability (Hartman-Grobman~\cite{GuckenheimerHolmes})}\, .
\end{eqnarray}
By {\em hyperbolicity} one typically  means in this context that a corresponding fixed point has the linearized dynamics that has no neutral (marginal) directions --- everything either decays or grows exponentially.
Consequently, the nonlinear RG flow near $\mathbf{g}^\ast$ is topologically equivalent to its linearization, and the classification into stable, unstable, and saddle points is robust under small perturbations of the beta function.
	
Non-hyperbolic fixed points satisfy $\det M=0$ and lie on bifurcation loci in coupling space. At such points the linear analysis breaks down, and higher-order terms in the beta function determine the local RG flow.

\subsection*{A2: Global structure and separatrix topology}

Beyond the local classification of fixed points, the RG flow possesses global structures that are invariant under smooth changes of coordinates on theory space. Of particular importance are the stable and unstable manifolds associated with fixed points, as these determine the topology of the flow and the partitioning of coupling space into distinct phases.
	
Let $\Phi_\ell:\mathcal{M}\rightarrow\mathcal{M}$ denote the RG flow generated by the beta-function vector field
\begin{eqnarray}
	\frac{d}{d\ell}\Phi_\ell(\mathbf{g}) \ = \ \beta(\Phi_\ell(\mathbf{g}))\, , \qquad \Phi_0=\mathrm{id}\, .
\end{eqnarray}
The global phase portrait is determined by invariant manifolds, namely the {\em stable} and {\em unstable manifolds} of a fixed point $\mathbf{g}^{\ast}$, which are defined as
\begin{eqnarray}
	&&W^s(\mathbf{g}^\ast) \ =  \ \{\mathbf{g}:\lim_{\ell\to\infty}\Phi_\ell(\mathbf{g}) \ = \ \mathbf{g}^\ast\}\, ,\nonumber \\[2mm]		&&W^u(\mathbf{g}^\ast) \ = \  \{\mathbf{g}:\lim_{\ell\to-\infty}\Phi_\ell(\mathbf{g}) \ = \ \mathbf{g}^\ast\}\, .
\end{eqnarray}
They generalize the notion of relevant and irrelevant directions beyond the linear approximation and
for a hyperbolic fixed point, these manifolds are invariant under the flow. In particular, for a saddle point they provide the boundaries separating trajectories with different infrared behavior.
	
For example, a critical fixed point with one relevant coupling and one irrelevant coupling (such as points $H_{\pm}$) has a one-dimensional stable manifold. On the other hand, for saddle fixed points (such as $I_{\pm}$) one has
\begin{eqnarray}
	\dim W^s \ + \  \dim W^u \ = \ \dim \mathcal{M}\, ,
\end{eqnarray}
and ensuing {\em separatrices} (or phase boundaries) are defined as
\begin{eqnarray}
	\Sigma \ = \  \bigcup_{\mathbf{g}^\ast \ \! \in \ \mathrm{saddle}} W^s(\mathbf{g}^\ast)\cup W^u(\mathbf{g}^\ast)\, .
\end{eqnarray}
So that $\Sigma$ represents a special trajectory (or collection of trajectories) that separates different flow behaviors.

The union of the stable and unstable manifolds of all saddle fixed points forms the {\em separatrix network}
\begin{eqnarray}
	\mathcal{M} \setminus \Sigma \ = \  \bigsqcup_\alpha \mathcal{B}_\alpha\, ,
\end{eqnarray}
which partitions coupling space into basins of attraction where each basin $\mathcal{B}_\alpha$ flows to a unique IR fixed point. The symbol $\bigsqcup$ reprsents a disjoint union.
	
Since diffeomorphic changes of coordinates map trajectories, fixed points, and invariant manifolds into one another, the separatrix structure and the associated phase-space topology are invariant properties of the RG flow.
	
\subsection*{A3: Topological invariants and homotopy structure}
	
Global, topological features of the RG flow that are invariant under smooth reparameterizations of theory space are also naturally encoded by homotopy classes of the beta-function vector field.
	
Let $\mathcal{M}$ denote the space of couplings and let
\begin{eqnarray}
	\beta:\;\;\; \mathcal{M}\ \rightarrow \ T\mathcal{M}\, ,
\end{eqnarray}
be the corresponding RG vector field. Two RG flows, represented by vector fields $\beta$ and $\beta'$, are said to be {\em homotopic} if there exists a continuous one-parameter family of vector fields
\begin{eqnarray}
	H_t:\;\;\; \mathcal{M}\ \rightarrow \ T\mathcal{M}\, ,
	\qquad
	t\in[0,1]\, ,
\end{eqnarray}
such that
\begin{eqnarray}
	H_0\ = \ \beta\, ,
   \qquad
	H_1 \ = \ \beta'\, .
\end{eqnarray}
Provided no fixed points are created, annihilated, or rendered non-hyperbolic during the deformation, the qualitative structure of the RG flow remains unchanged. In particular, the number and type of hyperbolic fixed points, together with the connectivity of their stable and unstable manifolds, are preserved.
	
A useful global invariant is obtained by compactifying coupling space and examining the behaviour of the flow at infinity. After Poincar\'{e} compactification (see Sec.~\ref{subsec:compact}), the boundary at infinity is represented by a sphere (or circle in two dimensions). Restricting the beta-function vector field to this boundary defines the normalised map
\begin{eqnarray}
	\hat{\beta}: \;\;\; S^1_{\infty}\rightarrow S^1\, ,
	\qquad
	\hat{\beta}
	\ = \
	\frac{\beta}{|\beta|}\, .
\end{eqnarray}
The associated winding number, or degree,
\begin{eqnarray}
	\deg(\hat{\beta})
	\ = \
	\frac{1}{2\pi}
	\oint_{S^1_{\infty}} d\theta
	\;\; \in \mathbb{Z}\, ,
\end{eqnarray}
measures the net rotation of the RG vector field along the boundary at infinity. Since the degree is a homotopy invariant, it cannot change under smooth coordinate transformations or field redefinitions.

The degree imposes global constraints on the admissible phase portrait of the RG flow. In particular, it is related through the Poincar\'{e}--Hopf theorem~\cite{Milnor1965Topology} to the sum of the indices of all fixed points in the compactified coupling space [see Eq.~(\ref{App.A.56.vb})]. Consequently, sources, sinks, and saddle points cannot be arbitrarily created or removed without altering the global topology of the flow. This provides a topological obstruction to transforming one RG phase portrait into another by a mere change of renormalisation scheme.

\section{Field Redefinitions, Coordinate Representations, and RG Scheme Dependence}
\label{sec:transformations}
%
%
In this appendix, we show how a field redefinition or a change of  RG scheme induces a reparametrization of theory space, thus altering the coordinate system, the operator basis, and the explicit form of the beta functions. We further demonstrate that such transformations cannot convert a genuinely unstable fixed point into a stable one in the exact theory.
	
We begin by noting that the Wilsonian RG defines a flow on the space of actions, or equivalently on the infinite-dimensional theory space generated by all local operators consistent with the symmetries of the problem~\cite{Wegner.74,Cardy1996book}. Upon choosing a basis of operators $\{\mathcal{O}_a\}$, the action can be expressed as
\begin{eqnarray}
	S[\phi;\mathbf{g}] \ = \  \sum_a g_a \OO_a[\phi]\, ,
\end{eqnarray}
where the couplings $g^a$ serve as coordinates on theory space.
	
In a given coordinate system, the RG flow is represented by beta functions
\begin{eqnarray}
	\frac{d g_a}{d \ell} \ = \  \beta_a(\mathbf{g})\, ,
	\qquad a\ = \ 1,2,\ldots M\, .
\end{eqnarray}
The beta functions themselves are not invariant objects. Rather, they are the components of a vector field on theory space expressed in a particular coordinate system.
	
In a Wilsonian implementation of the RG, one integrates out degrees of freedom within the momentum shell
\begin{eqnarray}
	\Lambda/b \ < \  |k|\  < \ \Lambda\, ,
\end{eqnarray}
and subsequently rescales momenta, fields, and couplings so as to restore the cutoff to its original value $\Lambda$, see, e.g.~\cite{Cardy1996book}.  This procedure depends on a number of choices, including the cutoff function, the blocking kernel, the field normalization, and the operator basis.  Different choices correspond to different RG schemes and alter the coordinate representation of the flow, and hence the explicit form of the beta functions, even though the underlying long-distance physics remains unchanged.
	
A particularly important class of such coordinate changes are field redefinitions~\cite{Kamefuchi.61,ZinnJustin}.
Consider a local, invertible, sufficiently regular change of variables 	$\phi = F[\widetilde\phi]$ in the functional integral
\begin{eqnarray}
	Z
	\ = \  \int \D[\phi]\, \ee^{-S[\phi]}
	\ = \  \int \D[\widetilde\phi]\,
	\det\!\left(\frac{\delta F}{\delta \widetilde\phi}\right)
	\ee^{-S[F[\widetilde\phi]]}\, .
\end{eqnarray}
Equivalently, the transformed action is~\cite{ZinnJustin,Kamefuchi.61}
\begin{eqnarray}
	S'[\widetilde\phi]
	\ = \ S[F[\widetilde\phi]]
	\ - \  \log \det\!\left(\frac{\delta F}{\delta \widetilde\phi}\right)\, .
\end{eqnarray}
The Jacobian is in general, regulator-dependent~\cite{ZinnJustin}.
For an infinitesimal field redefinition
\begin{eqnarray}
	\phi(x) \ = \  \widetilde\phi(x) \ + \  \alpha\,\Psi[\widetilde\phi](x)\, .
\end{eqnarray}
The transformed action is, to first order in $\alpha$
\begin{eqnarray}
	S'[\widetilde\phi]
	\ = \  S[\widetilde\phi]
	\ + \  \alpha \int \dd^d x\,
	\frac{\delta S}{\delta \widetilde\phi(x)}
	\Psi[\widetilde\phi](x)
	\ - \  \alpha\,\Tr\!\left(
	\frac{\delta \Psi}{\delta \widetilde\phi}
	\right)
	\ + \  \Order{\alpha^2}\, .
\end{eqnarray}
The induced operator is therefore redundant, since it merely parametrizes motion along an orbit generated by field redefinitions in theory space.
Redundant operators may appear in the action and in the beta functions, but they do not correspond to independent physical perturbations~\cite{Wegner.74}.
As an explicit example, consider the Euclidean $\phi^4$ theory
\begin{eqnarray}
	S[\phi]
	\ = \  \int d^d x
	\left[
	\frac12 (\partial_\mu \phi)^2
	\ + \  \frac12 m^2 \phi^2
	\ + \  \frac{g}{4!}\phi^4
	\right]\, .
\end{eqnarray}
Define the new field variable by
$\widetilde\phi(x) = \phi(x) + \epsilon \phi^3(x)$.
To first order in $\epsilon$, this is inverted as
\begin{eqnarray}
	\phi(x)
	\ = \ \widetilde\phi(x)
	\ - \  \epsilon \widetilde\phi^3(x)
	\ + \  \Order{\epsilon^2}\, .
\end{eqnarray}
The functional Jacobian is
\begin{eqnarray}
	J
	\ = \  \det\!\left(
	\frac{\delta \phi}{\delta \widetilde\phi}
	\right)
	\ = \  \det\!\left[
	\bigl(1 \ - \ 3\epsilon \widetilde\phi^2(x)\bigr)\delta(x-y)
	\right]\, ,
\end{eqnarray}
so that formally
\begin{eqnarray}
	\log J
	\ = \  -3\epsilon\,\delta_\Lambda(0)
	\int d^d x\,\widetilde\phi^2(x)
	\ + \  \Order{\epsilon^2}\, ,
\end{eqnarray}
where
\begin{eqnarray}
	\delta_\Lambda(0)
	\ \sim \ \int_{|p|<\Lambda} \frac{d^d p}{(2\pi)^d}\, ,
\end{eqnarray}
is a regulator-dependent contact term. It is useful to denote this local coefficient as
\begin{eqnarray}
	c_\Lambda \ \equiv \  \delta_\Lambda(0)\, .
\end{eqnarray}
Since
\begin{eqnarray}
	S'[\widetilde\phi] \ = \  S[\phi(\widetilde\phi)] \ - \  \log J\, ,
\end{eqnarray}
one obtains, to first order in $\epsilon$,
\begin{eqnarray}
	S'[\widetilde\phi] \ &=& \ \int d^d x
	\Bigg[{}
	\frac12 (\partial_\mu \widetilde\phi)^2 \ + \ \frac12 \bigl(m^2 \ + \  6\epsilon c_\Lambda\bigr)
	\widetilde\phi^2 \ + \ \left(\frac{g}{4!} - \epsilon m^2\right)
	\widetilde\phi^4 \nonumber
	\\[2mm]
	&& \ - \ \frac{\epsilon g}{6}\widetilde\phi^6
	\ - \ 3\epsilon\,
	\widetilde\phi^2(\partial_\mu\widetilde\phi)^2
	\Bigg] \ + \ \Order{\epsilon^2}\, .
\end{eqnarray}
Thus, the transformed action is not merely the original action with shifted $m^2$ and $g$. It also contains operators absent from the original two-coupling truncation, such as $\widetilde\phi^6$ and
$\widetilde\phi^2(\partial_\mu \widetilde\phi)^2$.
Therefore, the finite-dimensional subspace parametrized only by $(m^2,g)$ is not closed under this field redefinition. To describe the transformation faithfully, one must work in an enlarged theory space containing all generated operators~\cite{Wegner.74,ZinnJustin}. If one projects back onto the original truncated subspace, the result becomes scheme- and projection-dependent.
This example illustrates the general point that a local field redefinition changes the coordinates used to describe theory space and may generate redundant operators. It does not, by itself, change the physical RG trajectory in the full theory space \cite{Kamefuchi.61,Wegner.74,Cardy1996book}.
	
Let ${g}_a$ and ${g}'_{a}$ be two coordinate systems on theory space, related by a smooth, invertible, scale-independent map $g'_{a} = f_a(\mathbf{g})$.
The Jacobian of this coordinate change is
\begin{eqnarray}
	J_a{}_{b}(\mathbf{g}) \ = \  \frac{\partial f_a}{\partial g_b}\, .
\end{eqnarray}
The beta functions transform by the chain rule
\begin{eqnarray}
	\beta'_{a}(\mathbf{g}')
	\ = \  \frac{d g'_{a}}{d \ell}
	\ = \  \frac{\partial f_a}{\partial g_b}
	\frac{d g_b}{d \ell}
	\ = \  J_a{}_{b}(\mathbf{g})\,\beta_b(\mathbf{g})\, .
\end{eqnarray}
Thus, the beta functions transform as the components of a vector field under a change of coordinates on theory space.
	
If $\mathbf{g}^*$ is a fixed point in the original coordinates, i.e. $\beta_a(\mathbf{g}^*) = 0$, then the corresponding point $\mathbf{g}'^* = f(\mathbf{g}^*)$	is a fixed point in the new coordinates, because
\begin{eqnarray}
	\beta'_{a}(\mathbf{g}'^*)  \ = \ J_a{}_{b}(\mathbf{g}^*)\,\beta_b(\mathbf{g}^*) \ = \  0\, .
\end{eqnarray}
Therefore, a smooth invertible reparametrization of the full theory space maps fixed points to fixed points. It cannot create or annihilate fixed points, nor can it merge two distinct fixed points, unless the transformation is singular, non-invertible, or fails to cover the relevant region of theory space.
	
The linearized RG flow near a fixed point is governed by the stability matrix
\begin{eqnarray}
	M_a{}_{b}
	\ = \ \left. \frac{\partial \beta_a}{\partial g_b}
	\right|_{\mathbf{g}=\mathbf{g}^*}\, .
\end{eqnarray}
In the primed coordinates
\begin{eqnarray}
	M'_{a b} \ = \  \left.
	\frac{\partial \beta'_{a}}{\partial g'_{b}}
	\right|_{\mathbf{g}'=\mathbf{g}'_*}\, .
\end{eqnarray}
Using
\begin{eqnarray}
	\beta'_{a}(\mathbf{g}') \ = \  J_a{}_{c}(\mathbf{g})\,\beta_c(\mathbf{g})\, ,
\end{eqnarray}
and
\begin{eqnarray}
	\frac{\partial}{\partial g'_{b}}
	\ = \ (J^{-1})_d{}_{b}\frac{\partial}{\partial g_d}\, ,
\end{eqnarray}
one finds
\begin{eqnarray}
	M'_{ab}
	& = &
	\left. (J^{-1})_d{}_{b} \ \!
	\frac{\partial}{\partial g_d}
	\left[J_a{}_{c}(\mathbf{g})\,\beta_c(\mathbf{g})\right]
	\right|_{\mathbf{g}=\mathbf{g}_*} \nonumber \\[3mm]
	& = &
	\left. (J^{-1})_d{}_{b} \left[ \frac{\partial J_a{}_{c}}{\partial g_d}\,\beta_c(\mathbf{g})
	\ + \ J_a{}_{c}\frac{\partial \beta_c}{\partial g_d}
	\right] 	\right|_{\mathbf{g}=\mathbf{g}_*}\, .
\end{eqnarray}
At the fixed point, $\beta_c(\mathbf{g}^*)=0$, so the first term vanishes. Hence
\begin{eqnarray}
	M'_{ab}
	\ = \  J_a{}_{c}(\mathbf{g}^*)\,
	M_c{}_{d}\,
	(J^{-1})_d{}_{b}(\mathbf{g}^*)\, ,
\end{eqnarray}
or, in matrix notation,
\begin{eqnarray}
	M' \ = \  JMJ^{-1}\, .
\end{eqnarray}
A similarity transformation, therefore, relates the two stability matrices. Consequently, their eigenvalues are identical. The scaling dimensions associated with the linearized flow, and hence the critical exponents derived from them, are invariant under smooth, invertible, scale-independent reparametrizations of the full theory space~\cite{Wegner.74,Cardy1996book}.
The same statement applies to the number of relevant, irrelevant, and marginal directions. A relevant direction cannot be turned into an irrelevant direction by a regular change of variables. Likewise, a nonzero RG eigenvalue cannot be made zero by such a transformation. Marginal directions require separate analysis beyond linear order, but their existence as zero-eigenvalue directions of the linearized flow is also invariant under similarity transformations.
Thus, it is important to distinguish three different operations:
First, a field redefinition changes the field coordinates and usually induces a reparametrization of theory space together with redundant operators. Physical observables are unchanged, but the action and beta functions can look different.
Second, an RG scheme change, such as changing the cutoff profile, blocking prescription, subtraction convention, or normalization of fields and couplings, changes the coordinate representation of the RG vector field. In the exact, untruncated theory space, universal quantities such as critical exponents remain invariant \cite{Cardy1996book,ZinnJustin}.
Third, a truncation or projection of theory space is not a coordinate transformation. If one discards operators generated by a field redefinition or by the RG, then the projected beta functions need not be related by a similarity transformation. In such a truncated calculation, fixed-point locations, stability properties, and even the apparent existence of fixed points may become scheme dependent. Such changes are artifacts of the approximation, not genuine changes of the underlying RG flow~\cite{Wegner.74,Cardy1996book}.
If the coupling redefinition has explicit RG-time dependence, $g'_{a} = f_a(\mathbf{g},\ell)$, then the beta functions transform as
\begin{eqnarray}
	\beta'_{a}(\mathbf{g}',\ell)
	\ =\  \frac{\partial f_a}{\partial g_b}\,\beta_b(\mathbf{g})
	\ +\  \frac{\partial f_a}{\partial \ell}\, .
\end{eqnarray}
In this case, the simple vector-field transformation law is modified by the inhomogeneous term $\partial_\ell f_a$. Therefore, the standard invariance statements for fixed points and stability matrices require a scale-independent coordinate transformation, or else a formulation in which all variables have first been made autonomous, for example, by using dimensionless couplings and a fixed normalization convention.
Thus, under a smooth, invertible, scale-independent reparametrization of the full theory space, the RG beta functions transform as vector-field components, fixed points map one-to-one to fixed points, and the linearized stability matrices at corresponding fixed points are related by a similarity transformation. The critical exponents and the number of relevant, irrelevant, and marginal directions are therefore invariant. A field redefinition or RG scheme change can alter the coordinates, the operator basis, and the explicit beta functions, but it cannot convert a genuinely unstable fixed point into a stable one in the exact theory. Apparent changes of this kind signal either a singular transformation, explicit scale dependence, or a truncation/projection artifact.

\section{The rescaled set of RG equations}
\label{App_C}
	
As an explicit illustration of the concepts discussed in the previous two appendices, we present a stability analysis of the following RG flow equations
\begin{eqnarray}
	\frac{dx}{d\ell} \ &=&\ -\left(\frac{\epsilon}{2}\right)x \ + \ x^3 \ \equiv\ \beta_x\, ,
	\label{eq:orig1}\\[2mm]
	\frac{dy}{d\ell} \ &=&\ -\left(\frac{\epsilon}{2}\right)y \ + \ \frac{3}{2}y^3 \ \equiv\ \beta_y\, .
	\label{eq:orig2}
\end{eqnarray}
In particular, we will show that the flow topology associated with these equations differs qualitatively from that of the flow topology described by Eqs.~(\ref{eq:flowEquations})–(\ref{eq:A2}), so they cannot describe the same physical theory.

For completeness, we note that Eqs.~(\ref{eq:orig1})–(\ref{eq:orig2}) appear in the Supplemental Material of Ref.~\cite{YY-Chang.25}; the resulting RG flow portrait is depicted in Fig.~\ref{fig:RGflow}.

For $\epsilon>0$, one finds that there are nine fixed points:
\begin{eqnarray}
	(x^*,y^*) \ \in \
	\left\{	0,\ \pm\sqrt{\frac{\epsilon}{2}}
	\right\}
	\times	\left\{	0,\ \pm\sqrt{\frac{\epsilon}{3}}
	\right\}.
\end{eqnarray}
The vector field is completely decoupled. As a consequence, the stability matrix is diagonal:
\begin{eqnarray}
	M(x,y)	\ = \
	\begin{pmatrix}
		\partial_x \beta_x & \partial_y \beta_x \\[2mm]
		\partial_x \beta_y & \partial_y \beta_y
	\end{pmatrix}
	\ = \
	\begin{pmatrix}
		-\frac{\epsilon}{2}+3x^2 & 0 \\[2mm]
		0 & -\frac{\epsilon}{2}+\frac{9}{2} y^2
	\end{pmatrix}.
	\label{eq:stabilitymatrix}
\end{eqnarray}
Therefore, the RG eigenvalues at a fixed point $(x^*,y^*)$ are
\begin{eqnarray}
	\lambda_x \ = \ -\frac{\epsilon}{2} \ + \ 3(x^*)^2\, ,
	\qquad
	\lambda_y \ = \ -\frac{\epsilon}{2} \  + \ \frac{9}{2} (y^*)^2\, ,
\end{eqnarray}
and, for $\epsilon>0$, the complete stability classification is:\\
\begin{table}[h!]
	\centering
	\begin{tabular}{lccc}
		\toprule
		Fixed point & $\lambda_x$ & $\lambda_y$ & Type \\
		\midrule
		$\left(0,0\right)$
		& $-\epsilon/2$
		& $-\epsilon/2$
		& Stable (Gaussian)
		\\[1mm]
		$\left(\pm\sqrt{\epsilon/2},0\right)$
		& $\epsilon$
		& $-\epsilon/2$
		& Saddle
		\\[1mm]
		$\left(0,\pm\sqrt{\epsilon/3}\right)$
		& $-\epsilon/2$
		& $\epsilon$
		& Saddle
		\\[1mm]
$\left(\pm\sqrt{\epsilon/2},\pm\sqrt{\epsilon/3}\right)$
		& $\epsilon$
		& $\epsilon$
		& Repeller
		\\
		\bottomrule
	\end{tabular}
	\caption{Linear stability eigenvalues of the finite fixed points.}
	\label{tab:fixedpoints}
\end{table}
Thus, the Gaussian fixed point is the only locally stable finite fixed point for $\epsilon>0$.
The decoupled character of the RG flow equations also permits explicit solutions. For an equation
\begin{eqnarray}
	\frac{dz}{dt}\ =\ b z^3 \ - \ az\, ,
\end{eqnarray}
with $z(t=0)^2=z^2_0$, one finds
\begin{eqnarray}
	z(t)^2\ =\ \frac{a}{b-\left(b-\frac{a}{z_0^2}\right)e^{2at}}\, .
\end{eqnarray}
Comparison with the RG equations shows that $t=\ell$, $a=\epsilon/2$  and $b=1$ for $x$, while  for $y$ we have $b=3/2$.
For $\epsilon>0$, if $z^2_0<a/b$, then $z^2(\ell)\to0$. If $z^2_0>a/b$, the denominator reaches zero at finite $t$, signaling runaway to strong coupling.	
Poincare compactification refines the strong-coupling behavior by determining the asymptotic directions along which trajectories go to infinity.
Using polar variables $x=r\cos\theta,~y=r\sin\theta$ and rescaling $\ell$ by
\begin{eqnarray}
	\frac{d\tau}{d \ell} \ = \ r^2\, ,
\end{eqnarray}
yields for the angular flow
\begin{eqnarray}
	\frac{\dd\theta}{\dd\tau} \ =\
	\sin\theta\cos\theta
	\left(
	\frac{3}{2}\sin^2\theta \  - \ \cos^2\theta
	\right) \ \equiv \ F(\theta)\, .
	\label{eq:thetacompact}
\end{eqnarray}
Thus the compactified fixed directions $\theta^*$ are the $x$-axis directions ($\theta^*=0,\pi$), the $y$-axis directions ($\theta^*=\frac{\pi}{2},\frac{3\pi}{2}$), and the directions
$\frac{y}{x}=\pm\sqrt{\frac{2}{3}}$ ($\tan^2\theta^*=\frac{2}{3}$).
The sign of $F'(\theta^*)$ determines angular stability on the circle at infinity.
We find that the coordinate-axis directions are angularly attracting, while
the directions $y/x=\pm\sqrt{2/3}$ are angularly repelling.
	
Thus, the compactified fixed points at infinity are not finite-coupling RG fixed points. Rather, they classify asymptotic strong-coupling directions. Generic runaway trajectories approach one of the coordinate-axis infinities,so that either $|x|$ dominates $|y|$ or $|y|$ dominates $|x|$. The rays $y=\pm\sqrt{\frac{2}{3}}\,x$
are separatrices between strong-coupling sectors. They are unstable as angular directions.

Now compare Eqs.\! \eqref{eq:orig1}--\eqref{eq:orig2} with the coupled system
\begin{eqnarray}
	\frac{d g_1}{d \ell}
	\ &=& \
	-\frac{\epsilon}{2}g_1 \ + \ g_1^3\, ,
	\label{eq:coupled1}
	\\[2mm]
	\frac{d g_2}{d \ell}
	\ &=& \
	-\frac{\epsilon}{2}g_2
	\ + \ g_1^2 g_2
	\ + \ \frac32 g_2^3\, ,
	\label{eq:coupled2}
\end{eqnarray}
\textit{i.e.}, Eqs.~\eqref{eq:flowEquations}--\eqref{eq:A2}.
This coupled system has only five finite fixed points, \textit{i.e.}
\begin{eqnarray}
	(0,0)\, ,
	\qquad
	\left(0,\pm\sqrt{\frac{\epsilon}{3}}\right)\, ,
	\qquad
	\left(\pm\sqrt{\frac{\epsilon}{2}},0\right)\, .
\end{eqnarray}
By contrast, the system Eqs.\! \eqref{eq:orig1}--\eqref{eq:orig2} has nine finite fixed points. In particular, the four mixed fixed points
\begin{eqnarray}
	\left(\pm\sqrt{\frac{\epsilon}{2}},
	\pm\sqrt{\frac{\epsilon}{3}}\right)\, ,
\end{eqnarray}
of  Eqs.~\eqref{eq:orig1}--\eqref{eq:orig2} are absent from Eqs.\! \eqref{eq:flowEquations}--\eqref{eq:A2}.
	
A global smooth change of variables mapping one autonomous vector field to the other must map finite fixed points to finite fixed points and thus cannot change the number of finite fixed points. Since the system  \eqref{eq:orig1}--\eqref{eq:orig2} has nine finite fixed points for $\epsilon>0$, whereas the system \eqref{eq:flowEquations}--\eqref{eq:A2} has five, there is no global smooth diffeomorphism of the finite plane that maps the two systems into each other.

%
\section{Cancellation Mechanism and Planckian Dissipation}
\label{sec:appD}

{As discussed in Sec.~\ref{sec:omega-over-T}, observable-specific cancellations are not excluded on general RG grounds. In this appendix, we consider the proposal of Ref.~\cite{YY-Chang.25} and discuss to what extent the suggested cancellation mechanism is compatible with the scaling structure associated with a total-repeller quantum critical point.}

The U(1) FL$^*$, which corresponds to the strange-metal region in the RG phase diagram, is reached when both $\bar{g}$ and $J$ are irrelevant and flow to zero~\cite{YY-Chang.25}.
{According to Ref.~\cite{YY-Chang.25}, an intricate cancellation mechanism renders the transport scattering rate independent of $g$. If this mechanism is indeed operative, all bare couplings flowing to the U(1) FL$^*$ fixed point would display the same Planckian strange-metal behavior and would thereby evade the breakdown of $\omega/T$ scaling in the vicinity of the total-repeller QCP discussed in Sec.~\ref{sec:omega-over-T} \cite{private_Chung}.}

Here, we attempt to place the cancellation mechanism into the overall RG topology framework of critical charge Kondo breakdown developed in this article.

The cancellation mechanism should not be interpreted as ordinary scaling at a Gaussian fixed point.  At the Gaussian fixed point itself the interaction vertices vanish, $(g,J) \to (0,0)$, and the fixed-point theory is free, or at least asymptotically decoupled. If one sets $g=0$ directly in the low-energy action, there is no interaction process capable of producing a finite inelastic scattering rate.  A nonzero Planckian rate therefore cannot arise from the Gaussian fixed point in the usual sense.
	
The proposed cancellation has a more singular structure.  The same hopping/Kondo coupling $g$ that is irrelevant under the RG flow is also needed to generate the dynamics of the composite $\xi$-fermion.  Schematically, the inverse $\xi$-fermion propagator contains a generated kinetic coefficient
\begin{eqnarray}
	G_{\xi}^{-1}(\omega,k)
	\ \sim\
	\zeta\,(i\omega-\varepsilon_k) \ + \  \cdots\, ,
	\qquad
	\zeta \sim g^2 \, ,
\end{eqnarray}
while the corresponding $\xi$-fermion self-energy has the form
\begin{eqnarray}
	\Sigma_{\xi}''(\omega,T)
	\ \sim\
	g^2\,F(\omega,T)\, .
\end{eqnarray}
Thus, the effective scattering rate obtained from the dressed $\xi$-fermion propagator involves the ratio
\begin{eqnarray}
\Gamma_{\xi}(\omega,T)
\ \sim\
\frac{\Sigma_{\xi}''(\omega,T)}{\zeta}
\ \sim\
\frac{g^2 F(\omega,T)}{g^2}
\ \sim\
F(\omega,T)\, .
\end{eqnarray}
The explicit powers of $g^2$ cancel.  However, this cancellation is not a regular fixed-point statement: both numerator and denominator vanish as $g\to0$.  The finite result is obtained only if $g$ is kept nonzero while the $\xi$-fermion kinetic term and self-energy are generated, and only afterwards the ratio is formed.  Thus, the limits $g\to0$ and forming $\Sigma_{\xi}''/G_{\xi}^{-1}$ do not commute and the irrelevant coupling $g$ appears to be
dangerously irrelevant.  It is irrelevant for the RG flow of the fixed-point action, but it remains essential for the normalization and even the existence of the transport carrier whose scattering rate is being computed.  Therefore, if a universal Planckian rate emerges from this mechanism, it does not emerge from ordinary Gaussian scaling.  It emerges, if at all, from a singular cancellation controlled by the dangerously irrelevant coupling $g$.
	
This situation is analogous in spirit to the role of the quartic coupling in standard $\phi^4$ theory above the upper critical dimension.  For $d>4$, the Gaussian fixed point controls the critical exponents and the quartic coupling $u$ is RG-irrelevant.  Nevertheless, $u$ cannot simply be set to zero in the ordered phase, because it stabilizes the theory and fixes the magnitude of the order parameter, $\mathcal{F}[\phi]=r\phi^2+u\phi^4$, $\phi_0^2 \sim -r/u~(r<0)$.
The coupling $u$ is irrelevant at the fixed point, but physical quantities depend singularly on it.  This is the standard example of a dangerously irrelevant variable and is also the reason why naive hyperscaling fails above the upper critical dimension.
	
The proposed $\xi$-fermion cancellation has the same logical structure. The irrelevant coupling $g$ flows to zero, but the observable of interest is not obtained by evaluating the fixed-point theory at $g=0$.  Instead, it is obtained from a ratio in which both the generated kinetic term and the self-energy vanish with the same power of $g$.  Consequently, the cancellation may remove the explicit dependence on $g$, but it does not by itself establish ordinary hyperscaling or a conventional Gaussian fixed-point origin of the Planckian rate.

\bibliographystyle{unsrt}
\bibliography{references}

\end{document}